\documentclass[twocolumn,floatfix,eqsecnum,rmp]{revtex4}
\bibpunct{[}{]}{,}{n}{}{,}
\usepackage{graphicx}
\usepackage{amsmath}
\usepackage{bm}
\providecommand{\openone}{\leavevmode\hbox{\small1\kern-3.8pt\normalsize1}}
\begin{document}
%%%
\title{Electron-hole entanglement in the Fermi sea}
\author{C.W.J. Beenakker}
%\institute{
\affiliation{
Instituut-Lorentz, Universiteit Leiden,\\
P.O. Box 9506, 2300 RA Leiden, The Netherlands
}
%%%
%\begin{document}
\begin{abstract}
{\tt To be published in "Quantum Computers, Algorithms and Chaos",
International School of Physics Enrico Fermi, volume 162.}
\end{abstract}

\maketitle
\tableofcontents

\section{Introduction}
\label{intro}

\subsection{Preface}
\label{preface}

Enrico Fermi, to whom this School of Physics is dedicated, introduced the concept of the Fermi sea in a seminal 1926 paper \cite{Fer26}. At low temperature all energy levels in the conduction band of a metal are filled with electrons up to a maximal energy. This ``sea'' of electrons is called the Fermi sea and the maximal energy is the Fermi energy.\footnote{Historical note: Fermi surmised that these concepts would apply to bound electrons in an atom. The proper application to free electrons in a metal originated with Sommerfeld \cite{Bel94}.} 
An unfilled state in the Fermi sea behaves like a positively-charged electron, called a hole.

\begin{figure}
\includegraphics[width=0.95\linewidth]{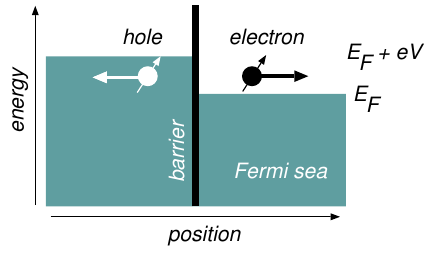}
\caption{Creation of an entangled electron-hole pair by application of a voltage difference $V$ between two metals separated by a tunnel barrier. The Fermi sea consists of the filled states in the conduction band. The spins of the electron ($e$) that has tunneled and the hole ($h$) it leaves behind are entangled in the state $2^{-1/2}(|\!\uparrow_{h}\uparrow_{e}\rangle+|\!\downarrow_{h}\downarrow_{e}\rangle)$.
}
\label{fig_barrier}
\end{figure}

Electron-hole excitations in the Fermi sea are created by application of a voltage $V$ over an insulating barrier, as illustrated in Fig.\ \ref{fig_barrier}. An electron can tunnel through the barrier, leaving behind a hole. The Fermi energy $E_{F}$ is higher by an amount $eV$ on one side of the barrier than on the other, so that the hole is elevated to the same energy as the electron.

A few years ago it occurred to us that the spatially separated electron-hole pair created by a tunneling event is a {\em spin-entangled Bell pair\/} \cite{Bee03}. This simple observation, that single-particle tunneling in a metal produces entanglement, came as a surprise. Earlier proposals for entanglers of spatially separated electrons (reviewed in Refs.\ \cite{Egu02,Rec04}) had relied on electron-electron interactions in one way or the other: the Coulomb interaction in a quantum dot \cite{Bur00,Oli02,Sar02}, the (phonon-mediated) pairing interaction in a superconductor \cite{Les01,Rec01,Ben02,Bou02,Sam03}, or the Kondo scattering by a magnetic impurity \cite{Cos01}.

Here we review the physics of entanglement in the Fermi sea from the three view points of production, detection, and utilization. The results presented in this review have all been published previously, except for the derivation in Sec.\ \ref{free_fermions} of the critical temperature above which the electron-hole entanglement vanishes.

In this introductory section we place our electron-hole entangler in a broader perspective by discussing similarities and differences with entanglers of excitons (Sec.\ \ref{excitons}) and of photons (Sec.\ \ref{optics_ana}).

\subsection{Exciton entanglers}
\label{excitons}

\begin{figure}
\includegraphics[width=0.6\linewidth]{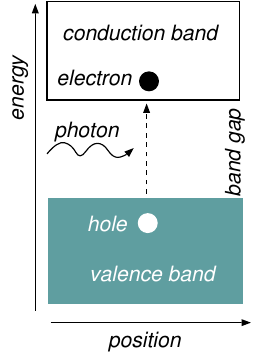}
\caption{Electron-hole pair (= exciton) formed by optical excitation of an electron from the (filled) valence band to the (empty) conduction band of a semiconductor. The key difference with the electron-hole pair in the Fermi sea of Fig.\ \ref{fig_barrier}, is that the latter can be produced without optical excitation (because electron and hole coexist at the same energy, both within the conduction band). 
}
\label{fig_exciton}
\end{figure}

In a semiconductor, holes can exist in the conduction band as well as in the valence band. A hole in the valence band bound to an electron in the conduction band is called an {\em exciton\/}, because it is produced by optical excitation of the electron across the band bap of the semiconductor (see Fig.\ \ref{fig_exciton}). In contrast, the electrons and holes in the conduction band of Fig.\ \ref{fig_barrier} coexist at the same energy, so that neither the creation nor the annihilation of the electron-hole pair involves any emission or absorption of a photon. The entanglement production and detection of electron-hole pairs in the Fermi sea is therefore purely electronic, without requiring any optical interface. 

Because the electron and hole that form an exciton are separated in energy by the band gap, their creation or annihilation is associated with photon emission or absorption. Optically induced entanglement of excitons in single and double quantum dots has been studied experimentally \cite{Che00,Len02} and theoretically \cite{Qui99,Hoh02,Shi04,Naz04,Che04}. In a single quantum dot the entanglement can be between the angular momentum of electron and hole forming a single exciton, but then the spatial separation of the electron-hole pair is problematic. In double quantum dots an exciton in one dot can become entangled with an exciton in the other dot, so the spatial separation is automatic.  

An altogether different theoretical proposal involves the ``energy-time'' entanglement of an electron in the conduction band with a hole in the valence band, excited at a p-n junction by a laser with a long coherence time \cite{Sca04}. In this form of entanglement (originating from quantum optics \cite{Fra89}) the energy and time of creation of each particle are uncertain, but the sum of the energies and the difference of the times are well-determined.

\subsection{Photon entanglers}
\label{optics_ana}

\begin{figure*}
\includegraphics[width=0.45\linewidth]{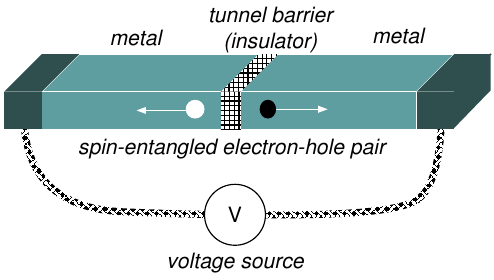}\qquad
\includegraphics[width=0.45\linewidth]{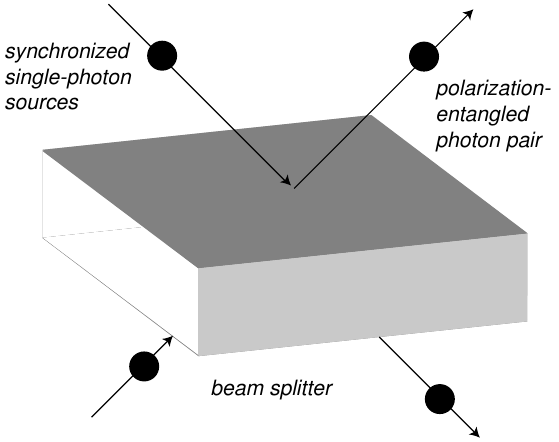}\\
\includegraphics[width=0.6\linewidth]{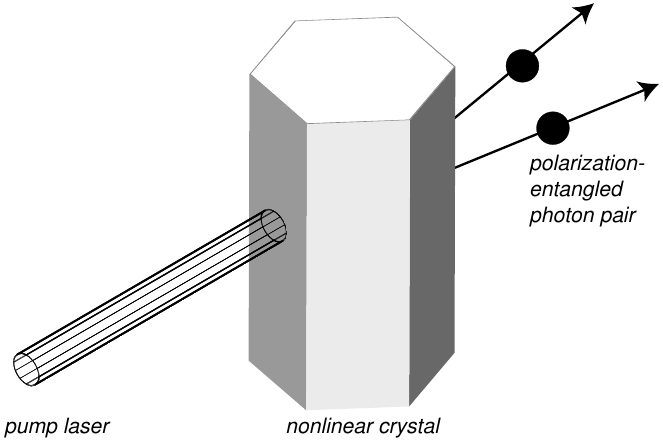}
\caption{Schematic illustration of the entanglement of electron-hole pairs by a tunnel barrier and of photons by a nonlinear crystal and a beam splitter. The differences and similarities of these entanglement mechanisms are discussed in the text. On the one hand, the Bell pair is produced spontaneously, without requiring synchronization at the source, by the voltage applied over the tunnel barrier and by the laser beam incident on the nonlinear crystal, while the beam splitter requires two synchronized single-photon sources. On the other hand, the tunnel barrier and the beam splitter both do not require interactions between the particles, while the nonlinear crystal requires photon-photon interactions to split the pump photon into two photons of lower frequency (down conversion).}
\label{fig_mechanisms}
\end{figure*}

The production and detection of entanglement is much further developed in optics than in electronics. It is therefore instructive to compare the mechanism for electron-hole entanglement of Fig.\ \ref{fig_barrier} with the two well-known mechanisms for photon entanglement illustrated in Fig.\ \ref{fig_mechanisms}.

The creation of a spin-entangled electron-hole pair by the applied voltage reminds one of the creation of a polarization-entangled photon pair by a pump laser in the process of spontaneous parametric down-conversion \cite{Man95}. The key difference is that the electronic process is governed by a single-particle Hamiltonian, while the optical process relies on photon-photon interactions in a nonlinear crystal (in order to split a pump photon of frequency $2\omega$ into two entangled photons of frequency $\omega$). What the two processes do have in common is that the Bell pair is produced at random times, triggered by vacuum fluctuations in the optical case and by stochastic tunnel events in the electronic case. The production rate is proportional to the nonlinear $\chi^{(2)}$ coefficient and the tunnel probability $\tau$, respectively. Although the time of creation of a Bell pair is random, the two particles are precisely synchronized in both processes --- without the need for any time resolution of the source.

If we seek an analogy in the realm of linear optics, one can think of the entanglement of two photons of opposite polarization that are simultaneously scattered by a 50/50 beam splitter \cite{Kni01,Sch01,Kim02,Xia02,Bos02}. With probability 1/2 the two photons end up in different arms in a polarization-entangled Bell state. There are two requirements for this process to succeed, which distinguish it from the electron-hole entangler: Firstly, noninteracting photons can not be entangled by a beam splitter if the sources are in local thermal equilibrium \cite{Xia02}. Somehow the Fermi sea, while being in local equilibrium, works around this optical no-go theorem. Secondly, the beam splitter relies on the indistinguishability of the two scattered photons, which is only guaranteed if the incoming pair of photons are scattered simultaneously. (More precisely, their wave packets should have a substantial overlap at the beam splitter \cite{Vel05}.) No such synchronization of the sources is required in the electronic case.

\section{Entanglement basics}
\label{ent_basics}

The concept of entanglement has been made precise and quantitative in the context of quantum information processing \cite{Nie00}. In this section we summarize what basic facts we need in order to calculate the amount of entanglement produced by the application of a voltage over a tunnel barrier. 

\subsection{Quantum versus classical correlations}
\label{q_vs_c_cor}

Loosely speaking, two spatially separated particles are entangled if their state can not be prepared from a product state by operating locally on each particle --- not even with the exchange of classical bits of information. The elementary entangled state is the spin singlet
\begin{equation}
|\Psi_{\rm Bell}\rangle=2^{-1/2}\bigl(|\!\uparrow\rangle_{A}|\!\downarrow\rangle_{B}-|\!\downarrow\rangle_{A}|\!\uparrow\rangle_{B}\bigr),\label{Belldef}
\end{equation}
also known as a Bell pair. The indices $A$, $B$ label the two particles and the arrows $\uparrow$, $\downarrow$ indicate two spin states. The state $|\Psi_{\rm Bell}\rangle$ can not be prepared locally starting from a product state such as $|\!\uparrow\rangle_{A}|\!\downarrow\rangle_{B}$.

The spins in the Bell state are correlated; a measurement of one spin, say with outcome $\uparrow$, projects the other spin on the opposite spin state $\downarrow$. The same applies to the mixed state with density matrix
\begin{equation}
\rho_{\rm mixed}=\tfrac{1}{2}\bigl(|\!\uparrow\rangle\langle\uparrow\! |\bigr)_{A}\otimes \bigl(|\!\downarrow\rangle\langle\downarrow\! |\bigr)_{B}+ \tfrac{1}{2}\bigl(|\!\downarrow\rangle\langle\downarrow\! |\bigr)_{A}\otimes \bigl(|\!\uparrow\rangle\langle\uparrow\! |\bigr)_{B}. \label{rhomixed}
\end{equation}
The difference between the quantum correlation of the pure entangled state (\ref{Belldef}) and the classical correlation of the mixed nonentangled state (\ref{rhomixed}) is that the quantum correlation persists if we measure the spin along a different axis, while the classical correlation is diminished. For example, the change of basis
\begin{equation}
|\!\uparrow\rangle\mapsto 2^{-1/2}\bigl(|\!\uparrow\rangle+|\!\downarrow\rangle\bigr),\;\;
|\!\downarrow\rangle\mapsto 2^{-1/2}\bigl(|\!\uparrow\rangle-|\!\downarrow\rangle\bigr) \label{rotatespin}
\end{equation}
(both for spin $A$ and spin $B$) leaves $|\Psi_{\rm Bell}\rangle$ invariant, but $\rho_{\rm mixed}$ becomes a mixture of parallel and anti-parallel spins.

\subsection{Bell inequality}
\label{Bell_ineq}

The Bell inequality is a test to distinguish quantum from classical correlations by comparing the correlation along different directions \cite{Bel64}. If spin $A$ is measured along unit vector $\bm{a}$ and spin $B$ along unit vector $\bm{b}$, then the correlator is the expectation value
\begin{equation}
C_{\bm{ab}}=\langle (\bm{a}\cdot\bm{\sigma})_{A}\otimes(\bm{b}\cdot\bm{\sigma})_{B}\rangle, \label{Cabdef}
\end{equation}
with $\bm{\sigma}=(\sigma_{x},\sigma_{y},\sigma_{z})$ the vector of Pauli matrices
\begin{equation}
\sigma_{x}=\begin{pmatrix}0&1\\ 1&0\end{pmatrix},\;\;
\sigma_{y}=\begin{pmatrix}0&-i\\ i&0\end{pmatrix},\;\;
\sigma_{z}=\begin{pmatrix}1&0\\ 0&-1\end{pmatrix}. \label{Paulidef}
\end{equation}

The Clauser-Horne-Shimony-Holt (CHSH) form of the Bell inequality reads \cite{Cla69}
\begin{equation}
{\cal B}=|C_{\bm{ab}}+C_{\bm{a'b}}+C_{\bm{ab'}}-C_{\bm{a'b'}}|\leq 2, \label{CHSHdef}
\end{equation}
for classically correlated spins. If ${\cal B}>2$ for some choice of unit vectors $\bm{a},\bm{b},\bm{a'}\,\bm{b'}$, then the spins are quantum correlated. The converse of this statement is not true: There exist mixed states that are entangled (in the sense that they can not be prepared locally) yet satisfy ${\cal B}\leq 2$ for all sets of unit vectors (see Sec.\ \ref{meas_mixed}).

An advantage of the CHSH form of the Bell inequality is that it can be used even if the detection efficiency is smaller than $1$. If a coincidence detection of spins $A$ and $B$ finds $N_{\pm}\in\{0,1\}$ spins pointing up or down at each detector, then the Bell-CHSH inequality reads
\begin{eqnarray}
{\cal B}&=&\left|\bigl\langle \bigl(N_{+}(\bm{a})-N_{-}(\bm{a})\bigr)\bigl(N_{+}(\bm{b})-N_{-}(\bm{b})\bigr)\bigr\rangle\right.\nonumber\\
&&\mbox{}+\bigl\langle \bigl(N_{+}(\bm{a'})-N_{-}(\bm{a'})\bigr)\bigl(N_{+}(\bm{b})-N_{-}(\bm{b})\bigr)\bigr\rangle\nonumber\\
&&\mbox{}+\bigl\langle \bigl(N_{+}(\bm{a})-N_{-}(\bm{a})\bigr)\bigl(N_{+}(\bm{b'})-N_{-}(\bm{b'})\bigr)\bigr\rangle\nonumber\\
&&\left.\mbox{}-\bigl\langle \bigl(N_{+}(\bm{a'})-N_{-}(\bm{a'})\bigr)\bigl(N_{+}(\bm{b'})-N_{-}(\bm{b'})\bigr)\bigr\rangle\right|\leq 2\nu,\nonumber\\
&&\label{CHSHnu}\\
\nu&=&\bigl\langle \bigl(N_{+}(\bm{a})+N_{-}(\bm{a})\bigr)\bigl(N_{+}(\bm{b})+N_{-}(\bm{b})\bigr)\bigr\rangle,\label{nudef}
\end{eqnarray}
where the detection efficiency $\nu$ is assumed to be independent of the unit vectors $\bm{a}$ and $\bm{b}$ along which spins $A$ and $B$ are measured.

\subsection{Entanglement measures for pure states}
\label{meas_pure}

It is rather straightforward to quantify the amount of entanglement present in a pure state, so we discuss this case first --- before proceeding to entanglement measures for mixed states in the next subsection.

Since an entangled state can not be prepared locally, one needs to exchange a certain amount of quantum information to create it out of a product state. This quantum information can take the form of Bell pairs, shared between $A$ and $B$. Bell pairs play the role of a ``currency'' (1 Bell pair = 1 bit of entanglement $\equiv$ 1 ebit), by means of which one can quantify entanglement. The average number of Bell pairs per copy needed to prepare a large number of copies of the pure state $|\Psi\rangle$ is given by
\begin{equation}
{\cal E}=-{\rm Tr}_{A}\,\rho_{A}\log_{2}\rho_{A},\;\;\rho_{A}={\rm Tr}_{B}\,|\Psi\rangle\langle\Psi|. \label{Edef}
\end{equation}
(The reduced density matrix $\rho_{A}$ of subsystem $A$ is obtained by tracing out the degrees of freedom of subsystem $B$.) 

The quantity ${\cal E}$ is called the entanglement entropy, or entanglement of formation. It can be used to quantify the amount of entanglement present in the pure state of any system with a Hilbert space ${\cal H}={\cal H}_{A}\otimes {\cal H}_{B}$ that can be divided into two subsystems $A$ and $B$ (a socalled bipartite system). It is important to emphasize that entanglement entropy is defined relative to a particular partitioning. Indeed, a system might well be separable relative to ${\cal H}={\cal H}_{A}\otimes {\cal H}_{B}$ and entangled relative to ${\cal H}={\cal H}'_{A}\otimes {\cal H}'_{B}$.

In the special case that each subsystem contains a single qubit, one can use an equivalent measure of entanglement called the concurrence. A normalized two-qubit state $|\Psi\rangle$ has the form
\begin{equation}
|\Psi\rangle=\sum_{i=1}^{2}\sum_{j=1}^{2}\gamma_{ij}|i\rangle_{A}|j\rangle_{B},\;\;{\rm Tr}\,\gamma\gamma^{\dagger}=1, \label{Psigeneral}
\end{equation}
where we have labeled $1\equiv\uparrow$, $2\equiv\downarrow$. Substitution into Eq.\ (\ref{Edef}) gives the following expression for the entanglement entropy in terms of the $2\times 2$ matrix of coefficients $\gamma$:
\begin{equation}
{\cal E}={\cal F}\left(\tfrac{1}{2}+\tfrac{1}{2}\sqrt{1-4\,{\rm Det}\,\gamma\gamma^{\dagger}}\right), \label{ECrelation}
\end{equation}
where the function ${\cal F}(x)$ is defined by
\begin{equation}
{\cal F}(x)=-x\log_{2} x-(1-x)\log_{2}(1-x).\label{calFdef}
\end{equation}
The quantity
\begin{equation}
{\cal C}=2\sqrt{{\rm Det}\,\gamma\gamma^{\dagger}}\label{Cdef}
\end{equation}
is called the concurrence of the two qubits. It is in one-to-one relation with ${\cal E}$ but has a somewhat simpler expression. A Bell pair has unit concurrence and carries one bit of entanglement, while both ${\cal C}$ and ${\cal E}$ vanish for a product state.

From an experimental point of view, the Bell inequality is the most direct way to quantify the degree of entanglement of two qubits. By maximizing the correlator ${\cal B}$ in Eq.\ (\ref{CHSHdef}) over the unit vectors one measures the Bell parameter
\begin{equation}
{\cal B}_{\rm max}=\max_{\bm{a},\bm{b},\bm{a'}\,\bm{b'}}{\cal B}.\label{Bmaxdef}
\end{equation}
The relation between ${\cal B}_{\rm max}$ and ${\cal C}$ is \cite{Gis91}
\begin{equation}
{\cal B}_{\rm max}=2\sqrt{1+{\cal C}^{2}}.\label{BmaxCrelation}
\end{equation}
The Bell parameter varies from $2$ to $2\sqrt{2}$ when the concurrence varies from $0$ to $1$. As mentioned in the previous subsection, any entangled pure state violates the Bell inequality (${\cal B}_{\rm max}>2$ if and only if ${\cal C}>0$).

\subsection{Entanglement measures for mixed states}
\label{meas_mixed}

The density matrix $\rho$ of a mixed state can be decomposed into pure states $|\Psi_{n}\rangle$ with positive weight $p_{n}$,
\begin{equation}
\rho=\sum_{n}p_{n}|\Psi_{n}\rangle\langle\Psi_{n}|,\;\;
p_{n}>0,\;\;\sum_{n}p_{n}=1.\label{rhodecomposed}
\end{equation}
The states $|\Psi_{n}\rangle$ are normalized to unity, $\langle\Psi_{n}|\Psi_{n}\rangle=1$, but they need not be orthogonal. The convex-sum decomposition (\ref{rhodecomposed}) is therefore not unique --- there are many equivalent representations of $\rho$ as a mixture of pure states.

A mixed state in the bipartite Hilbert space ${\cal H}_{A}\otimes{\cal H}_{B}$ is nonentangled (= separable) if there exists a convex-sum decomposition (\ref{rhodecomposed}) into pure product states \cite{Wer89}, meaning that $|\Psi_{n}\rangle=|\Phi_{n}\rangle_{A}|\Phi'_{n}\rangle_{B}$ with $|\Phi_{n}\rangle_{A}\in{\cal H}_{A}$ and $|\Phi'_{n}\rangle_{B}\in{\cal H}_{B}$ for all $n$. The entanglement entropy ${\cal E}$ of a separable mixed state vanishes, with the definition
\begin{equation}
{\cal E}=\min_{\{\Psi_{n},p_{n}\}}\,\sum_{n}p_{n}{\cal E}(\Psi_{n}). \label{Emixedstatedef}
\end{equation}
Here ${\cal E}(\Psi_{n})$ is the entanglement entropy of the pure state $|\Psi_{n}\rangle$, defined in Eq.\ (\ref{Edef}), and the minimum is taken over all convex-sum decompositions (\ref{rhodecomposed}) of $\rho$.

The entanglement entropy, or entanglement of formation, of a mixed state is an upper bound to the average number of Bell pairs per copy that it costs to prepare many copies $\rho\otimes\rho\otimes\rho\otimes\cdots$ of the state $\rho$ \cite{Ben96a}. This socalled entanglement cost would be equal to ${\cal E}$ if ${\cal E}$ would be an additive quantity, meaning that ${\cal E}(\rho\otimes\rho)=2{\cal E}(\rho)$. The additivity of the entanglement entropy has been proven for certain classes of mixed states \cite{Vid02}, but not in general. It might be that for certain mixed states entanglement has a ``discount'', in the sense that the cost per copy is less for many copies than for a single copy.

The entanglement entropy is also an upper bound to the average number of Bell pairs per copy that one can extract (or ``distill'') from many copies of an entangled state, using only local operations and classical communication \cite{Ben96b}. The production of entangled pure states out of Bell pairs is reversible, meaning that ${\cal E}$ equals the distillable entanglement. For mixed states this is not the case in general: The distillable entanglement can be less than ${\cal E}$, so a fraction of the Bell pairs can be lost in the conversion from Bell pairs to entangled mixed states and back to Bell pairs.

The calculation of ${\cal E}$ for a general multi-qubit bipartite mixed state is more complicated than for a pure state, because of the need to minimize over convex-sum decompositions. In the case of two qubits a closed form expression for ${\cal E}$ exists, due to Wootters \cite{Woo98}:
\begin{eqnarray}
&&{\cal E}={\cal F}\left(\tfrac{1}{2}+\tfrac{1}{2}\sqrt{1-{\cal C}^{2}}\right),\label{ECrelation2}\\
&&{\cal C}=\max\{0,\sqrt{\lambda_{1}}-\sqrt{\lambda_{2}}-\sqrt{\lambda_{3}}-\sqrt{\lambda_{4}}\}.\label{Cdef2}
\end{eqnarray}
The $\lambda_{i}$'s are the eigenvalues of the matrix product
\[
\rho\cdot(\sigma_{y}\otimes\sigma_{y})\cdot\rho^{\ast}\cdot(\sigma_{y}\otimes\sigma_{y}), 
\]
in the order $\lambda_{1}\geq\lambda_{2}\geq\lambda_{3}\geq\lambda_{4}$. The quantity ${\cal C}$ is again called the concurrence. For a pure state (with $\rho^{2}=\rho$) the definition (\ref{Cdef2}) of ${\cal C}$ is equivalent to Eq.\ (\ref{Cdef}).

As a simple example we take the rotationally invariant mixed state
\begin{equation}
\rho_{\rm Werner}(\xi)=\tfrac{1}{4}(1-\xi)\openone_{4}+\xi|\Psi_{\rm Bell}\rangle\langle\Psi_{\rm Bell}|,\;\; -\tfrac{1}{3}\leq \xi\leq 1,\label{rhoWerner}
\end{equation}
with $\openone_{4}$ the $4\times 4$ unit matrix. This is the socalled Werner state \cite{Wer89}. The concurrence is
\begin{equation}
{\cal C}=\max\{0,{\tfrac{1}{2}}(3\xi-1)\}.\label{CWerner}
\end{equation}
The Werner state is separable for $|\xi|\leq 1/3$. Notice that separable mixed states, unlike separable pure states, occupy a region in parameter space of nonzero measure.

For a mixed state there is no one-to-one relation between the Bell parameter (\ref{Bmaxdef}), which quantifies the maximal violation of the CHSH inequality (\ref{CHSHdef}), on the one hand, and the entanglement entropy or concurrence, on the other hand. Depending on the density matrix, ${\cal B}_{\rm max}$ can take on values between $2{\cal C}\sqrt{2}$ and $2\sqrt{1+{\cal C}^{2}}$. The general formula \cite{Hor95,Ver02}
\begin{equation}
{\cal B}_{\rm max}=2\sqrt{u_{1}+u_{2}}\label{Bmixed}
\end{equation}
for the dependence of ${\cal B}_{\rm max}$ on $\rho$ involves the two largest eigenvalues $u_{1},u_{2}$ of the real symmetric $3\times 3$ matrix $R^{T}R$ constructed from $R_{kl}={\rm Tr}\,\rho\,\sigma_{k}\otimes\sigma_{l}$. In the case of the Werner state (\ref{rhoWerner}) one finds ${\cal B}_{\rm max}=2\xi\sqrt{2}$, so this state is entangled without violating the CHSH inequality if $1/3<\xi\leq 1/\sqrt{2}$.

\subsection{Particle conservation}
\label{c_cons}

As discussed in Sec.\ \ref{q_vs_c_cor}, to distinguish classical from quantum mechanical correlations it is essential that the strength of the correlation is compared in different bases. In other words, to detect entanglement between two separated qubits one needs to be able to rotate each qubit individually, creating superpositions like Eq.\ (\ref{rotatespin}). As emphasized by Wiseman and Vaccaro \cite{Wis03b}, if a conservation law prevents the rotation, then the entanglement reduces effectively to a classical correlation --- becoming inaccessible as a quantum resource.

For electrons the restriction on the accessible entanglement imposed by conservation of particles is of primary importance. Particle conservation prevents the local creation of a state that is in a superposition of a different number of particles: If $|0\rangle$ denotes an empty single-electron state and $|1\rangle$ denotes a filled state, then the superposition
\begin{equation}
|0\rangle\mapsto 2^{-1/2}\bigl(|0\rangle+|1\rangle\bigr),\;\;
|1\rangle\mapsto 2^{-1/2}\bigl(|0\rangle-|1\rangle\bigr) \label{rotatecharge}
\end{equation}
can not be created locally. As a consequence, the single-electron state
\begin{equation}
|\Psi_{1}\rangle=2^{-1/2}\bigl(|0\rangle_{A}|1\rangle_{B}-|1\rangle_{A}|0\rangle_{B}\bigr) \label{Psi1}
\end{equation}
has ${\cal E}=1$, like the two-particle Bell state (\ref{Belldef}), but this entanglement is inaccessible because of particle conservation. (Such useless  entanglement has been dubbed ``fluffy bunny'' entanglement \cite{Wis03a}.)

To account for particle conservation in a bipartite multi-electron state we need to project the density matrix $\rho$ onto sectors ${\cal N}_{pq}$ of Fock space with an integer particle number $p$ at $A$ and $q$ at $B$. This projection eliminates unobservable coherences between sectors of different particle number. The total accessible entanglement entropy ${\cal E}_{\rm part}$, constrained by particle conservation, is the weighted sum of the entanglement from the individual sectors \cite{Wis03b},
\begin{eqnarray}
&&{\cal E}_{\rm part}=\sum_{p,q}w_{pq}{\cal E}(\rho_{pq}),\label{Ewsum}\\
&&\rho_{pq}=\frac{1}{w_{pq}}\Pi_{pq}^{\vphantom{\dagger}}\rho\Pi_{pq}^{\dagger},\;\;w_{pq}={\rm Tr}\,\Pi_{pq}^{\vphantom{\dagger}}\rho\Pi_{pq}^{\dagger}.\label{wpqdef}
\end{eqnarray}
Here $\Pi_{pq}$ is the projection operator onto ${\cal N}_{pq}$.

For the state (\ref{Psi1}) we have two sectors $p=0,q=1$ and $p=1,q=0$, with $w_{01}=w_{10}=1/2$ and $\rho_{01}=(|0\rangle\langle 0|)_{A}(|1\rangle\langle 1|)_{B}$, $\rho_{10}=(|1\rangle\langle 1|)_{A}(|0\rangle\langle 0|)_{B}$. The accessible entanglement
\begin{equation}
{\cal E}_{\rm part}=\tfrac{1}{2}{\cal E}(\rho_{01})+\tfrac{1}{2}{\cal E}(\rho_{10})=0 \label{Epart1}
\end{equation}
vanishes, as expected, because $\rho_{01}$ and $\rho_{10}$ are product states.

\subsection{Phase reference}

The restriction on the accessible entanglement imposed by particle conservation can be (partially) removed if the two parties $A$ and $B$ share a phase reference \cite{Kit04,Enk05,Bar05}. In the quantum optical context, a phase reference is an identical pair of coherent states (one at $A$ and one at $B$). In the electronic context of interest here, the most natural candidate for a phase reference is a pair of superconductors with the same phase of the order parameter (which we may therefore set equal to zero). Andreev reflection at the superconductor converts a state with $N$ electrons into a state with $N\pm 2$ electrons, adding or extracting a Cooper pair to or from the superconducting condensate. By combining Andreev reflection and normal reflection one can realize the superpositions
\begin{equation}
|0\rangle\mapsto 2^{-1/2}\bigl(|0\rangle+|2\rangle\bigr),\;\;
|2\rangle\mapsto 2^{-1/2}\bigl(|0\rangle-|2\rangle\bigr). \label{rotatecharge2}
\end{equation}
This is indeed how the charge qubit in a socalled ``Cooper pair box'' is rotated \cite{Nak99}.

The rotation (\ref{rotatecharge}), involving a charge difference of $1e$ rather than $2e$, remains inaccessible, in view of the lack of a macroscopically coherent state of unpaired electrons.

\section{How to entangle free particles}
\label{howto_entangle}

One way to entangle particles is by letting them interact with each other. If the Hamiltonian $H=H_{A}+H_{B}+H_{AB}$ of two spatially separated particles $A$ and $B$ contains an interaction term $H_{AB}$, then their time dependent state $e^{-iHt/\hbar}|\Psi\rangle_{A}|\Psi\rangle_{B}$ will evolve from a product state into an entangled state. Since the particles are spatially separated, and hence distinguishable, it is irrelevant whether they are fermions or bosons.

Indistinguishable particles may become entangled by scattering from an external potential, such as the beam splitter or tunnel barrier of Fig.\ \ref{fig_mechanisms}, even if they do not interact with each other. A fundamental difference now appears between fermions and bosons, as we discuss in the following subsections.

\subsection{Free bosons}
\label{free_bosons}

Whether or not a beam splitter is able to entangle noninteracting particles depends on the quantum state of the sources \cite{Kni01,Sch01,Kim02,Bos02,Xia02}. {\em Free bosons can not be entangled if the sources are in (local) thermal equilibrium.\/} For a proof of this ``no-go theorem'' we follow Xiang-bin \cite{Xia02}.

\begin{figure}
\includegraphics[width=0.6\linewidth]{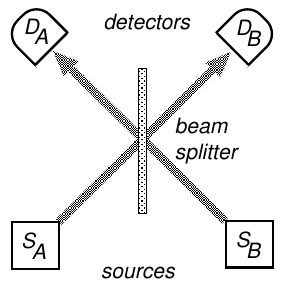}
\caption{Schematic setup for entanglement of indistinguishable particles by a beam splitter. Two sources $S_{A}$ and $S_{B}$ create a separable initial state. The beam splitter scatters the particles into two detectors $D_{A}$ and $D_{B}$. An entangled final state may result if the detectors can not distinguish from which source a particle originated. For bosons no entanglement can be produced by the beam splitter if the sources are in thermal equilibrium. For fermions this ``no-go theorem'' does not apply.}
\label{fig_beamsplitter}
\end{figure}

We consider the geometry of Fig.\ \ref{fig_beamsplitter}. The density matrix $\rho_{\rm in}$ of the multi-mode state incident from the sources onto the beam splitter can be written in the coherent state representation,
\begin{eqnarray}
&&\rho_{\rm in}=\int d\bm{\alpha}\, P(\bm{\alpha})|\bm{\alpha}\rangle\langle\bm{\alpha}|, \label{rhoinPalpha}\\
&&|\bm{\alpha}\rangle=\exp\left(\bm{a}^{\dagger}\cdot\bm{\alpha}-\bm{a}\cdot\bm{\alpha}^{\ast}\right)|0\rangle. \label{alphadef}
\end{eqnarray}
The coherent state $|\bm{\alpha}\rangle=|\alpha_{1},\alpha_{2},\ldots\rangle$ is an eigenstate of the annihilation operator $a_{n}$ with complex eigenvalue $\alpha_{n}$. We have abbreviated $d\bm{\alpha}\equiv\prod_{n}d{\rm Re}\,\alpha_{n}d{\rm Im}\,\alpha_{n}$. 

The real function $P(\bm{\alpha})$ may take on negative values. In that case the state $\rho_{\rm in}$ is called ``nonclassical'', because $P(\bm{\alpha})$ can not be interpreted as a classical probability distribution \cite{Man95}. A thermal state is a classical state. (It has a Gaussian $P(\bm{\alpha})\propto\exp(-\sum_{n}|\alpha_{n}|^{2}/f_{n})$, with $f_{n}$ the Bose-Einstein occupation number of mode $n$.) Nonclassical states include photon-number (or Fock) states and squeezed states.

The beam splitter transforms the annihilation operators $\bm{a}=\{a_{1},a_{2},\ldots\}$ of the incoming state into annihilation operators $\bm{b}=\{b_{1},b_{2},\ldots\}$ of the outgoing state. This is a unitary transformation,
\begin{equation}
b_{n}=\sum_{m}S_{nm}a_{m},\;\;S\cdot S^{\dagger}=\openone, \label{bSa}
\end{equation}
defined by the scattering matrix $S$. The density matrix $\rho_{\rm out}$ of the outgoing state is obtained from $\rho_{\rm in}$ with the help of Eq.\ (\ref{bSa}). The result is
\begin{eqnarray}
&&\rho_{\rm out}=\int d\bm{\beta}\,P(S^{\dagger}\cdot\bm{\beta})|\bm{\beta}\rangle\langle\bm{\beta}|, \label{rhooutPbeta}\\
&&|\bm{\beta}\rangle=\exp\left(\bm{b}^{\dagger}\cdot\bm{\beta}-\bm{b}\cdot\bm{\beta}^{\ast}\right)|0\rangle. \label{betadef}
\end{eqnarray}
We have made a change of variables from $\bm{\alpha}$ to $\bm{\beta}=S\cdot\bm{\alpha}$ and used that $d\bm{\alpha}=d\bm{\beta}$ because of the unitarity of $S$.

We now group the outgoing modes into $N_{A}$ modes that are detected by $A$ and $N_{B}$ modes that are detected by $B$. Since the creation and annihilation operators of different modes commute, we may write $|\bm{\beta}\rangle$ as a product state,
\begin{equation}
|\bm{\beta}\rangle=\left(\prod_{n=1}^{N_{A}}e^{b_{n}^{\dagger}\beta_{n}-b_{n}\beta_{n}^{\ast}}\right)\left(\prod_{n=N_{A}+1}^{N_{A}+N_{B}}e^{b_{n}^{\dagger}\beta_{n}-b_{n}\beta_{n}^{\ast}}\right)|0\rangle. \label{betaproduct}
\end{equation}
The pure state $|\bm{\beta}\rangle$ is therefore separable for all $\bm{\beta}$. Referring to the definition of a separable mixed state in Sec.\ \ref{meas_mixed}, we conclude that $\rho_{\rm out}$ in Eq.\ (\ref{rhooutPbeta}) is separable if $P\geq 0$ for all $\bm{\beta}$, because in that case the density matrix has a decomposition into separable pure states with positive weights.

Nonclassicality of the sources is therefore a necessary condition for entanglement by single-particle scattering \cite{Xia02}. This condition is violated by a thermal source, proving the no-go theorem.

\subsection{Free fermions}
\label{free_fermions}

In contrast to bosons, {\em fermions can be entangled by single-particle scattering even if the sources are in (local) thermal equilibrium\/} \cite{Bee03}. The no-go theorem of the previous subsection does not apply, because it relies on a distinction between classical and nonclassical bosonic density matrices without a fermionic analogue.\footnote{The fermionic analogue of the coherent state decomposition (\ref{rhoinPalpha}) involves an integration over Grassman variables, rather than over complex numbers \cite{Cah99}. A Grassman integral is not a valid convex-sum decomposition of a density matrix, because it lacks a positive integration measure.}
To demonstrate the free-fermion entanglement in its simplest form, we consider a single-mode,\footnote{The single-mode restriction is not essential in the case of spin-independent scattering; in the eigenbasis of the transmission matrix product $tt^{\dagger}$, multiple modes represent multiple ``copies'' of the spin-entangled state, so the total entanglement production is simply the sum of the contributions from the individual modes.} 
spin-degenerate conductor at zero temperature, containing a barrier with spin-independent tunnel probability $\tau$.

We refer to the geometry of Fig.\ \ref{fig_barrier}. Tunnel attempts in the energy range $eV$ occur with a rate $eV/h$, independently for spin-up and spin-down electrons \cite{Lev93}. In each time interval $h/eV$, either 0, 1, or 2 electron-hole pairs are created, transforming the unperturbed Fermi sea $|0\rangle$ into the superposition\footnote{Tunneling creates a coherent superposition rather than an incoherent mixture because of the indistinguishability of the electrons.}
\begin{eqnarray}
|0\rangle &\mapsto& (1-\tau)|0\rangle-e^{2i\phi}\tau|\!\uparrow\downarrow\rangle_{h}|\!\uparrow\downarrow\rangle_{e}\nonumber\\
&&\mbox{}-e^{i\phi}\sqrt{2\tau(1-\tau)}\underbrace{2^{-1/2}\left(|\!\uparrow\rangle_{h}|\!\uparrow\rangle_{e}+|\!\downarrow\rangle_{h}|\!\downarrow\rangle_{e}\right)}_{\mbox{Bell pair}},\nonumber\\
&&\label{0mapsto}
\end{eqnarray}
with $\phi$ the phase difference of the (spin-independent) transmission and reflection amplitudes. In each product of kets $|\cdot\rangle$, the first ket refers to the hole excitations at the left of the barrier and the second ket refers to the electron excitations at the right of the barrier. The third term (labeled ``Bell pair'') involves the creation of a single electron-hole pair (with spin up or spin down), while the first and second term involve the creation of 0 and 2 electron-hole pairs, respectively.

As discussed in Sec.\ \ref{c_cons}, the total entanglement ${\cal E}_{\rm part}$, constrained by particle conservation, is the sum of the entanglement carried by the 0, 1, and 2-particle states. This gives $(1-\tau)^{2}\times 0+2\tau(1-\tau)\times 1+\tau^{2}\times 0=2\tau(1-\tau)$ bits. To obtain the entanglement production in a detection time $t_{\rm det}$ we should multiply by the number $t_{\rm det}\times eV/h$ of tunnel attempts in that time interval, resulting in
\begin{equation}
{\cal E}_{\rm part}/t_{\rm det}=\frac{2eV}{h}\tau(1-\tau).\label{E0mapsto}
\end{equation}
The maximum entanglement production rate of $eV/2h$ is reached for $\tau=1/2$ (corresponding to a 50/50 beam splitter). In the tunneling limit $\tau\ll 1$ one has ${\cal E}_{\rm part}/t_{\rm det}=2eV\tau/h$.

Notice that the states at the left-hand-side and at the right-hand-side of Eq.\ (\ref{0mapsto}) are related by a single-particle unitary transformation, and yet the left-hand-side is separable while the right-hand-side is entangled. There is no contradiction because entanglement is only invariant under {\em local\/} unitary transformations. Scattering by the tunnel barrier mixes the degrees of freedom at the left and at the right, so it is a nonlocal transformation --- which has the capability to transform a separable state into an entangled state.

The result (\ref{E0mapsto}) points to an important connection between entanglement production and shot noise: The entanglement production rate in the case of spin-independent scattering is proportional to the spectral density $P_{\rm noise}$ of the current fluctuations through the conductor, given by \cite{Les89,But90}
\begin{equation}
P_{\rm noise}=2eV\frac{2e^{2}}{h}\tau(1-\tau)=2e^{2}{\cal E}_{\rm part}/t_{\rm det}.\label{shotnoisepower}
\end{equation}
We will return to this connection in Sec.\ \ref{epump}.

In Ref.\ \cite{Bee03} the more general case of spin-dependent scattering was analyzed (still for a single-mode conductor and at zero temperature). The calculation is reviewed in Appendix \ref{app_spindep}. The transmission matrix $t$ of the tunnel barrier is a $2\times 2$ complex matrix. The transmission eigenvalues $\tau_{1},\tau_{2}\in[0,1]$ are the eigenvalues of the Hermitian matrix product $tt^{\dagger}$. The result for the entanglement production rate is

\begin{widetext}
\begin{eqnarray}
{\cal E}_{\rm part}/t_{\rm det}&=&\frac{eV}{h}(\tau_{1}+\tau_{2}-2\tau_{1}\tau_{2}){\cal F}\left(\tfrac{1}{2}+\tfrac{1}{2}\sqrt{1-{\cal C}^{2}}\right)\nonumber\\
&=&\frac{eV}{h}\bigl\{(\tau_{1}+\tau_{2}-2\tau_{1}\tau_{2})\log_{2}(\tau_{1}+\tau_{2}-2\tau_{1}\tau_{2})\nonumber\\
&&\mbox{}-\tau_{1} (1 - \tau_{2}) \log_{2}[\tau_{1} (1 - \tau_{2})] - \tau_{2}(1 - \tau_{1})\log_{2}[\tau_{2}(1 - \tau_{1})]\bigr\}\label{ECrelation3}\\
{\cal C}&=&\frac{2\sqrt{\tau_{1}\tau_{2}(1-\tau_{1})(1-\tau_{2})}}{\tau_{1}+\tau_{2}-2\tau_{1}\tau_{2}}.\label{Cdef4}
\end{eqnarray}
\end{widetext}
The concurrence (\ref{Cdef4}) of the electron-hole pair reaches its maximum value ${\cal C}=1$ for the spin-independent case $\tau_{1}=\tau_{2}\equiv\tau$. In that case Eq.\ (\ref{ECrelation3}) reduces to Eq.\ (\ref{E0mapsto}).

Note that the concurrence is entirely determined by the transmission eigenvalues; the eigenvectors of $tt^{\dagger}$ do not contribute. This means, in particular, that channel mixing does not degrade the entanglement as long as the transmission eigenvalues remain unaffected. Note also that the direct relation (\ref{shotnoisepower}) between entanglement production rate and shot noise power no longer exists in the case of spin-dependent scattering. The shot noise power in this case equals $P_{\rm noise}=2eV(e^{2}/h)[\tau_{1}(1-\tau_{1})+\tau_{2}(1-\tau_{2})]$, which is not in one-to-one relationship with Eq.\ (\ref{ECrelation3}).

\begin{figure}
\includegraphics[width=0.8\linewidth]{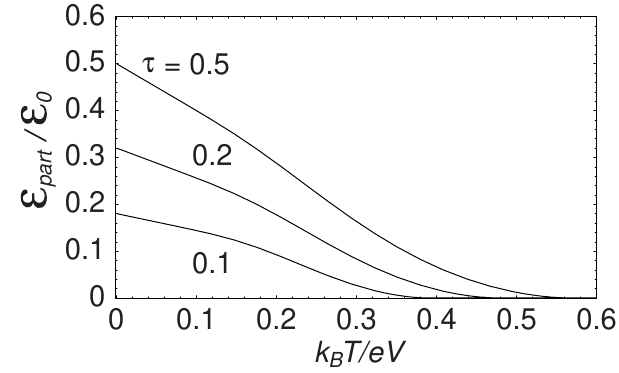}
\caption{Temperature dependence of the entanglement production ${\cal E}_{\rm part}$ (in units of ${\cal E}_{0}=eVt_{\rm det}/h$), calculated from Eq.\ (\ref{Eparttotal}). Results are shown for three values of the spin-independent transmission probability $\tau$. (The same plots are obtained if $\tau$ is replaced by $1-\tau$.)}
\label{fig_Epartplot}
\end{figure}

\begin{figure}
\includegraphics[width=0.8\linewidth]{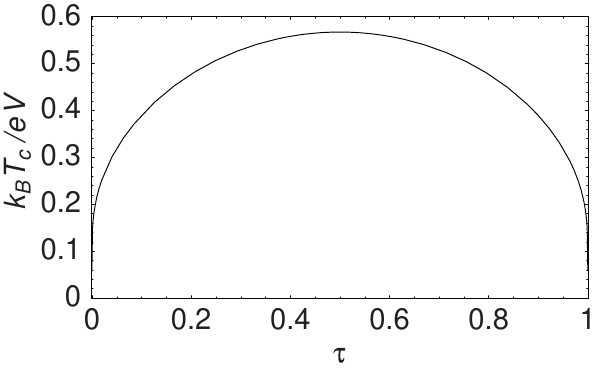}
\caption{The entanglement production of Fig.\ \ref{fig_Epartplot} is identically zero for $T\geq T_{c}$, with the critical temperature given by Eq.\ (\ref{Tceq}) and plotted in this figure.}
\label{fig_T_crit}
\end{figure}

So far we have assumed zero temperature. At nonzero temperature the electronic state is mixed rather than pure, and the rate of entanglement production decreases. General considerations \cite{Fin05} require that the entanglement vanishes identically for temperatures $T$ greater than a critical temperature $T_{c}$. The calculation of $T_{c}$ is presented in Appendix \ref{app_finiteT}, for the case of spin-independent scattering.  The result (\ref{Eparttotal}) is plotted in Fig.\ \ref{fig_Epartplot}. We find that $T_{c}$ is given by
\begin{equation}
\tau(1-\tau)\sinh^{2}(eV/2k_{B}T_{c})=1/4,\label{Tceq}
\end{equation}
as plotted in Fig.\ \ref{fig_T_crit}. When $T_{c}$ is approached from below, the entanglement vanishes with a logarithmic singularity,
\begin{equation}
{\cal E}_{\rm part}\propto(T_{c}-T)^{2}\ln(T_{c}-T)^{-1},\;\;{\rm when}\;\;T\uparrow T_{c}.\label{logsing}
\end{equation}
The maximal critical temperature for a given voltage is reached for $\tau=1/2$, when $T_{c}=0.57\,eV/k_{B}$. In the tunneling regime one has\footnote{
A second critical temperature $T'_{c}$ exists, above which the Bell-CHSH inequality can not be violated. In general $T'_{c}\leq T_{c}$. In the tunneling regime $\tau\ll 1$ the two critical temperatures coincide.}
\begin{equation}
T_{c}=\frac{eV}{k_{B}\ln(1/\tau)},\;\;{\rm if}\;\;\tau\ll 1.\label{Tcsmalltau}
\end{equation}
Because the dependence of $T_{c}$ on $\tau$ is only logarithmic, the critical temperature will be of order $eV/k_{B}$ even for very small tunnel probabilities.

\begin{figure}
\includegraphics[width=0.8\linewidth]{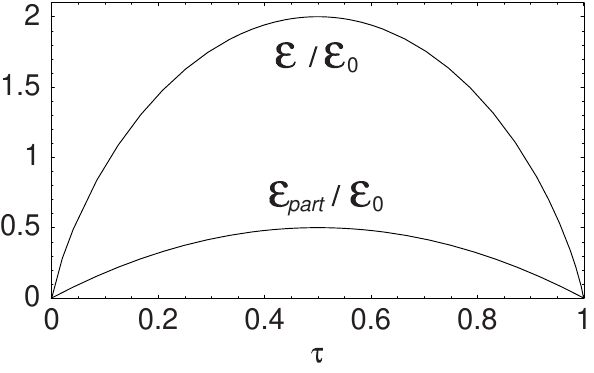}
\caption{Comparison of the unconstrained entanglement production ${\cal E}$ [given by Eq.\ (\ref{Emodes1})] with the entanglement production ${\cal E}_{\rm part}$ constrained by particle conservation [given by Eq.\ (\ref{ECrelation3})]. Both quantities are normalized by ${\cal E}_{0}=eVt_{\rm det}/h$ and plotted as a function of the spin-independent transmission probability $\tau_{1}=\tau_{2}\equiv\tau$. Zero temperature is assumed.}
\label{fig_emodespart}
\end{figure}

We close this subsection by calculating how much the constraint of particle number conservation reduces the entanglement production ${\cal E}_{\rm part}$ below the unconstrained value ${\cal E}$. We take zero temperature, so we can use the pure-state formula (\ref{Edef}) for ${\cal E}$. The result is\footnote{
The entanglement production rate given without further discussion in footnote 24 of Ref.\ \cite{Bee03} is this unconstrained value ${\cal E}$, rather than the more appropriate constrained value ${\cal E}_{\rm part}$.}
\begin{equation}
{\cal E}/t_{\rm det}=\frac{eV}{h}[{\cal F}(\tau_{1})+{\cal F}(\tau_{2})],\label{Emodes1}
\end{equation}
which in the tunnel limit simplifies to
\begin{equation}
{\cal E}/t_{\rm det}=\frac{eV}{h}(-\tau_{1}\log_{2}\tau_{1}-\tau_{2}\log\tau_{2}),\;\;{\rm if}\;\;\tau_{1},\tau_{2}\ll 1.\label{Emodes2}
\end{equation}
The constrained entanglement production (\ref{ECrelation3}) simplifies in the tunnel limit to
\begin{eqnarray}
{\cal E}_{\rm part}/t_{\rm det}&=&\frac{eV}{h}[-\tau_{1}\log_{2}\tau_{1}-\tau_{2}\log\tau_{2}\nonumber\\
&&\mbox{}+(\tau_{1}+\tau_{2})\log_{2}(\tau_{1}+\tau_{2})],\;\;{\rm if}\;\;\tau_{1},\tau_{2}\ll 1.\nonumber\\
&&\label{Eparticles}
\end{eqnarray}
We conclude that the constraint of particle conservation reduces the entanglement production rate by an amount $(eV/h)(\tau_{1}+\tau_{2})\log_{2}(\tau_{1}+\tau_{2})$ in the tunnel limit. The two quantities ${\cal E}$ and ${\cal E}_{\rm part}$ are compared in Fig.\ \ref{fig_emodespart} for the case of spin-independent transmission probability.

\section{Spin versus orbital entanglement}
\label{spinvsorbit}

\begin{figure*}
(a)\!\!\!\!\!\!\!\!\includegraphics[width=
0.45\linewidth]{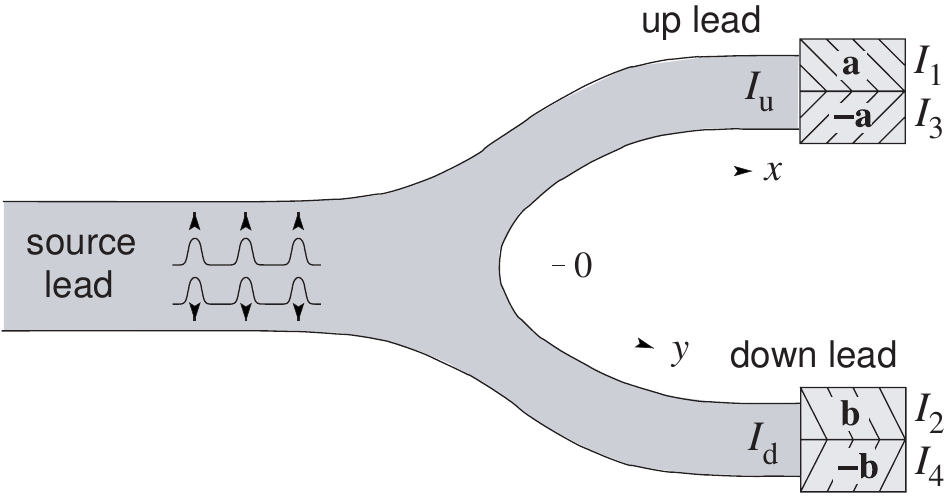}\quad
\includegraphics[width=
0.45\linewidth]{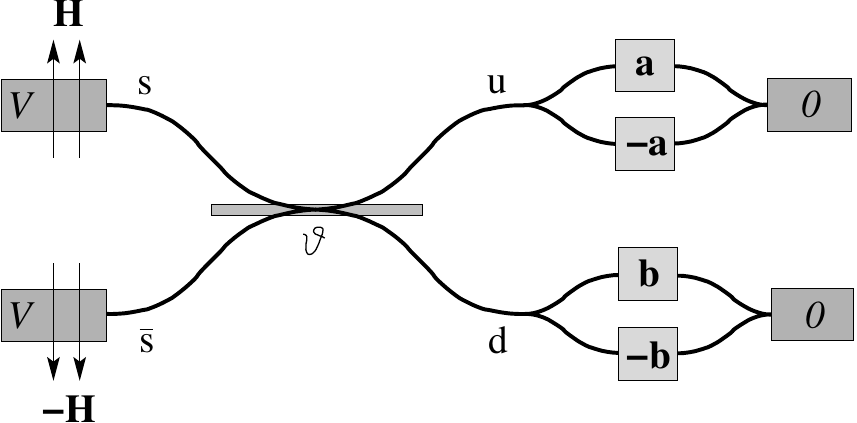}\!\!\!\!\!\!(b)\bigskip\\
\centerline{(c)\includegraphics[width=0.4\linewidth]{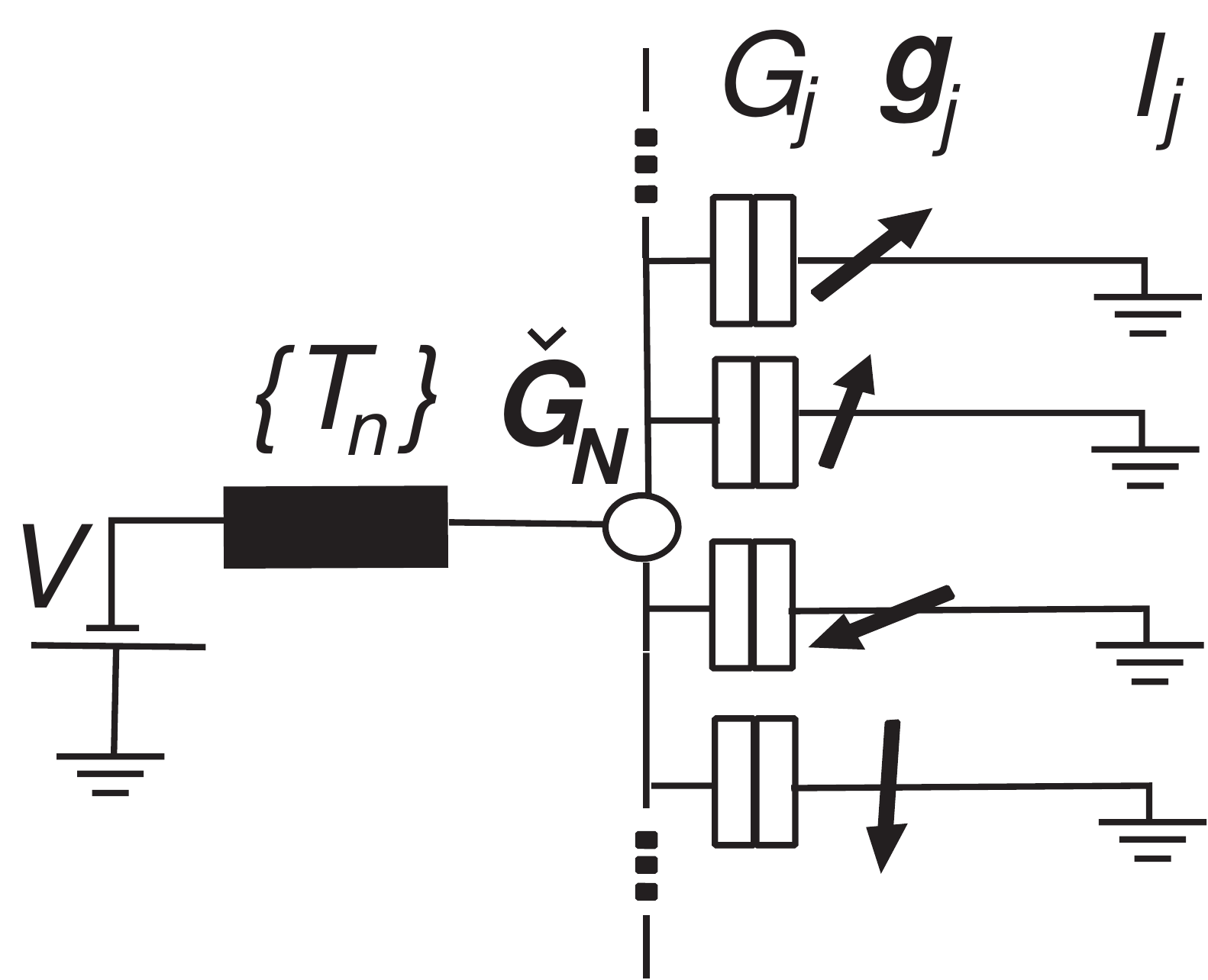}}
\caption{Three geometries to produce and detect spin entanglement in a normal conductor with ferromagnetic contacts. Panels a, b, and c are taken, respectively, from Refs.\ \cite{Leb05}, \cite{Leb04}, and \cite{Lor05}. The magnetization axis in the contacts is indicated by the vectors $\bm{a},\bm{b}$ or by arrows.}
\label{fig_geometries_spin}
\end{figure*}

\begin{figure*}
(a)\includegraphics[width=0.6\linewidth]{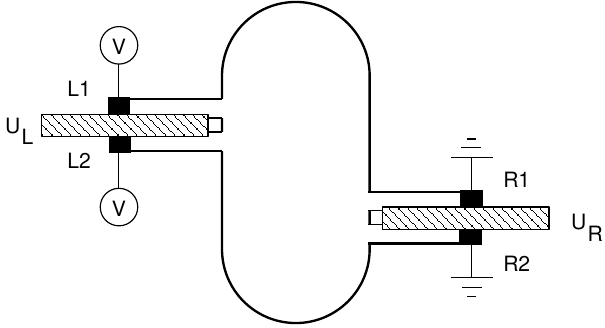}\\
(b)\includegraphics[width=0.6\linewidth]{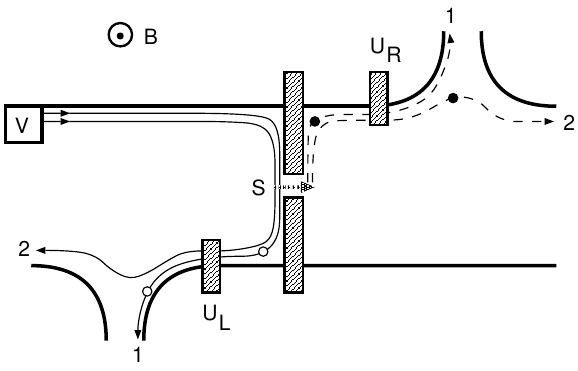}\\
(c)\includegraphics[width=0.5\linewidth]{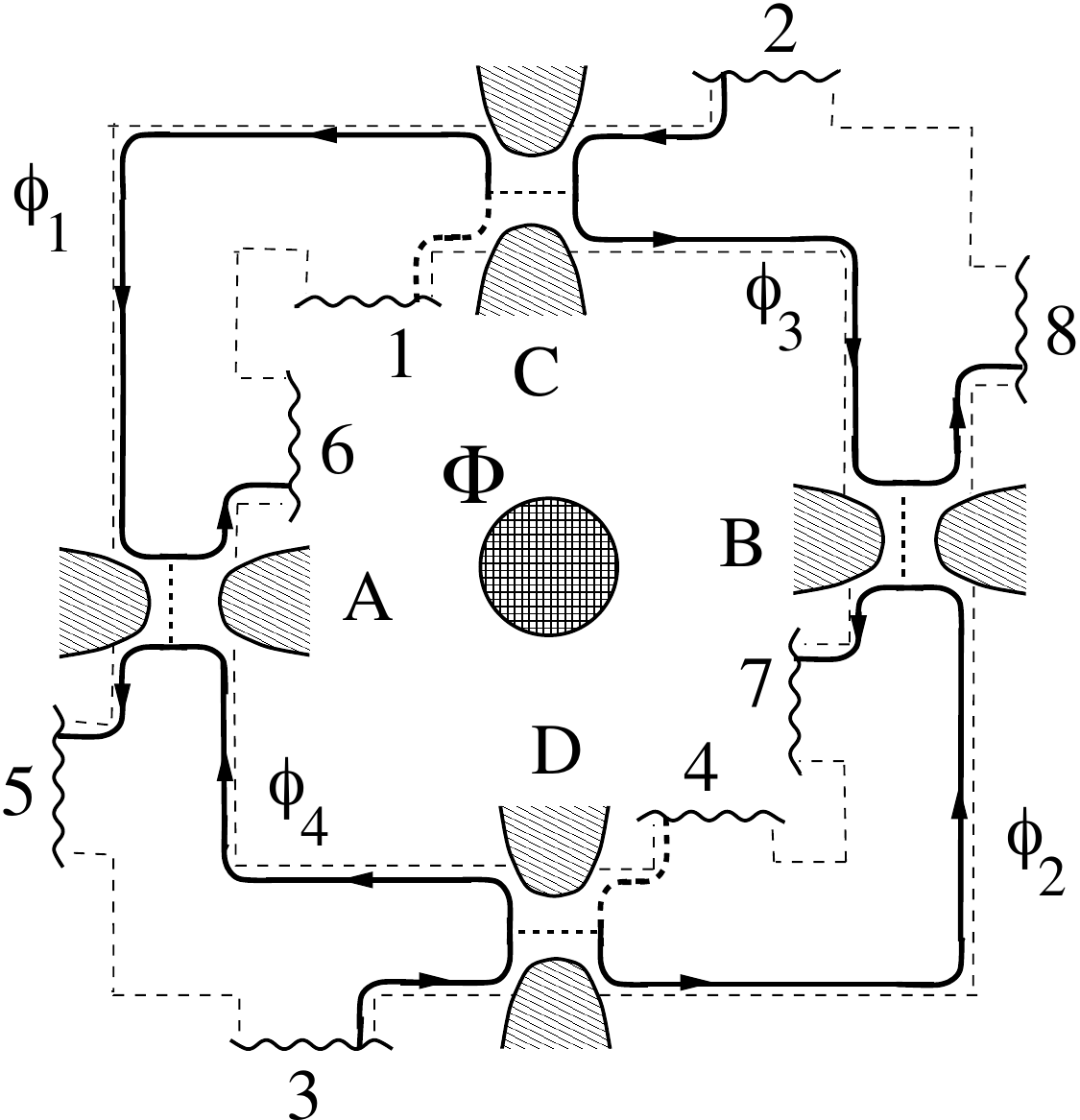}
\caption{Three geometries to produce and detect orbital entanglement in zero magnetic field (a) and in the quantum Hall effect regime (b,c). Panels a, b, and c are taken, respectively, from Refs.\ \cite{Bee04a}, \cite{Bee03}, and \cite{Sam04}. The solid (dashed) lines in panels b,c indicate filled (empty) edge channels, with the arrows pointing in the direction of motion.}
\label{fig_geometries_orbit}
\end{figure*}

Any normal-metal conductor containing a localized scatterer can be used to
entangle the outgoing states to the left and right of the scatterer. The
particular implementation described originally in Ref.\ \cite{Bee03} uses tunneling between edge channels in the integer quantum Hall effect. A variety of other implementations have since been proposed, as shown in Figs.\ \ref{fig_geometries_spin} and \ref{fig_geometries_orbit}. 

In all implementations a stream of entangled electron-hole pairs is produced by application of a voltage over a normal-metal conductor. What differs is the degree of freedom which is entangled and the means by which the entanglement is detected. The three implementations from Refs.\ \cite{Leb05,Leb04,Lor05} shown in Fig.\ \ref{fig_geometries_spin} entangle the spin degree of freedom and use ferromagnetic contacts to measure the spin-resolved correlator needed to test for violation of the Bell inequality (as proposed by Kawabata \cite{Kaw01} and Chtchelkatchev et al.\ \cite{Cht02}). The three implementations from Refs.\ \cite{Bee03,Bee04a,Sam04} shown in Fig.\ \ref{fig_geometries_orbit} entangle a spatial (= orbital) degree of freedom instead of spin and use non-ferromagnetic contacts for the detection (as proposed by Samuelsson, Sukhorukov, and B\"{u}ttiker \cite{Sam03}).

The implementations of Fig.\ \ref{fig_geometries_spin} have the advantage that a spin-entangled electron-hole pair is much less sensitive to decoherence than an orbitally entangled pair, simply because most external degrees of freedom couple to the position of the electron rather than to its spin. The entanglement production is optimal because, assuming that the scattering by the normal-metal conductor is spin-independent, each electron-hole pair represents a maximally entangled Bell pair (= spin singlet). The maximum entanglement production rate of $eV/2h$ bits/second is reached for a 50/50 beam splitter [cf.\ Eq.\ (\ref{E0mapsto})]. There is no need for the conductor to be single-mode --- multiple orbital modes simply produce multiple copies of the spin-entangled Bell pair. The disadvantage of these implementations is that it is difficult to resolve the different spin components of the current. A ferromagnetic contact with an adjustable magnetization axis may provide the required spin filtering capability \cite{Gij97}. In order to avoid backscattering, the conductance of the non-ferromagnetic conductor should be much less than the total conductance of the ferromagnetic contacts \cite{Lor05}. An alternative non-ferromagnetic spin filter might be provided by a quantum dot in a strong magnetic field \cite{Rec00,Han04}.

The orbital entanglement in Figs.\ \ref{fig_geometries_orbit}a,b,c refers to different spatial degrees of freedom. The quantum dot of Fig.\ \ref{fig_geometries_orbit}a has two single-channel point contacts at the left (labeled $L_{1},L_{2}$) and two more at the right (labeled $R_{1},R_{2}$). Each point contact plays the role of one of the spin directions in the spin entanglers. In the Hall bar geometry of Fig.\ \ref{fig_geometries_orbit}b this role is played by the two quantum Hall edge channels (labeled by their Landau level index 1,2) at the left and right of the constriction. There is only a single Landau level in the disk geometry of Fig.\ \ref{fig_geometries_orbit}c --- the role of the spin direction being played by the source contact 2 or 3 from which an electron reaches the detectors at A and B.

The quantum dot entangler of Fig.\ \ref{fig_geometries_orbit}a has the advantage that no magnetic field is required for its operation, but the disadvantage that backscattering at the detector (back into the quantum dot) can not easily be avoided. A strong perpendicular magnetic field, as in Figs.\ \ref{fig_geometries_orbit}b,c, suppresses backscattering: The Lorentz force constrains the electrons to move unidirectionally along equipotentials at the edge of the two-dimensional electron gas. (For a tutorial on edge channel transport, see Ref.\ \cite{Bee91}.) Let us discuss these two edge channel entanglers in some more detail.

The Hall bar entangler contains a beam splitter formed by a split gate electrode (shaded rectangles at the center of Fig.\ \ref{fig_geometries_orbit}b), as realized experimentally in Ref.\ \cite{Hen99}. The current in edge channel 1 can be measured independently of that in edge channel 2 by diverting an equipotential into a point contact, as in Ref.\ \cite{Wee89}. This capability to filter edge channels is not enough for a test of the Bell inequality; the filter should also have an adjustable eigenbasis. In the spin filters of Fig.\ \ref{fig_geometries_spin} the eigenbasis is just the magnetization axis, which can be rotated by a magnetic field. Equivalently, one could rotate the spin at fixed magnetization axis of the spin filter. The edge channel filters have a fixed eigenbasis, so one would need to ``rotate'' or mix the edge channels before they are filtered. The two gate electrodes labeled $U_{L}$ and $U_{R}$ in Fig.\ \ref{fig_geometries_orbit}b should serve this purpose, by creating a strong local electric field that causes inter-edge-channel transitions. Such edge channel mixers have not yet been demonstrated in the laboratory, however. 

The electron-hole pairs produced in this geometry have a concurrence given by Eq.\ (\ref{Cdef4}). For an optimal entanglement the two transmission eigenvalues $\tau_{1}$ and $\tau_{2}$ should not be widely different. This is problematic, since edge channel number 1 tunnels over a shorter distance through the constriction than edge channel number 2, and therefore $\tau_{2}$ will in general be $\ll\tau_{1}$. It may be possible to resolve this problem by introducing disorder into the constriction, in order to mix the edge channels while they are tunneling and equalize the transmission eigenvalues.

Samuelsson, Sukhorukov, and B\"{u}ttiker \cite{Sam04} found a way to avoid the need to mix edge channels from different Landau levels by working with a single Landau level in a disk geometry (Fig.\ \ref{fig_geometries_orbit}c). This geometry is the electronic analogue of the Hanbury-Brown-Twiss interferometer in quantum optics \cite{Man95}. Edge channels circulate counter-clockwise along the outer edge of the disc and clockwise along the inner edge. Four split-gate electrodes (labeled A,B,C,D) couple the inner and outer edge. Four electrical contacts (numbered 2,3,5,8) are connected to the outer edge and four more (numbered 1,4,6,7) are connected to the inner edge. All contacts are grounded, except contacts 2 and 3 at voltage $V$. This number 2,3 of the source contact replaces the Landau level index 1,2 in the Hall bar geometry. The split gates A and B replace the edge channel mixers $U_{L}$ and $U_{R}$, while the two split gates C and D together replace the beam splitter at the center of the Hall bar. By adjusting the gates C and D separately one can readily equalize the two transmission eigenvalues and ensure that the electron-hole pair arriving at gates A and B is maximally entangled. Tunneling through split gates A and B provides the mixing needed to test the Bell inequality. Such mixers within a single Landau level have been realized experimentally \cite{Ji03}.

\section{Entanglement detection by noise measurements}
\label{detection}

\subsection{Tunneling regime}
\label{detection_tunneling}

We emphasized in our presentation of the electron-hole entangler, in Sec.\ \ref{intro}, that no synchronization at the sources is required: The electron and hole are automatically produced at the same instant in time, since they originate from a single tunnel event. We did not yet address the issue whether synchronization might be required at the detectors.

In a typical optical experiment \cite{Man95}, the two photon counters that test for entanglement of a photon pair are synchronized by a coincidence circuit. The time resolution required to detect the current pulses from individual tunnel events is difficult to reach in solid state electronics. Fortunately, it is not needed. As pointed out by Samuelsson, Sukhorukov, and B\"{u}ttiker \cite{Sam03}, low-frequency noise measurements can test for the violation of the Bell inequality even if individual current pulses are not resolved.

\begin{figure}
\includegraphics[width=0.8\linewidth]{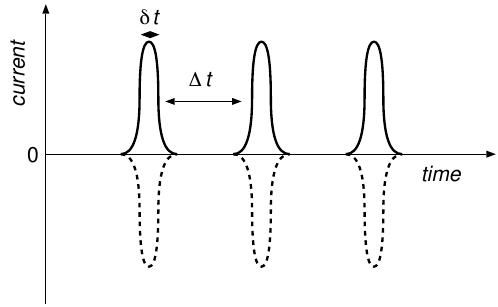}
\caption{Current pulses of electrons (solid) and holes (dashed) produced at a tunnel barrier with transmission probability $\tau\ll 1$. The width $\delta t$ of the pulses is smaller than their spacing $\Delta t$ by a factor $\tau$. Low-frequency noise measurements correlate the spin of each electron with that of every hole, but only the spins from coincident current pulses contribute (because the spins from non-coincident pulses are uncorrelated). In this way the electron-hole entanglement can be detected without requiring time-resolved measurements.}
\label{fig_pulses}
\end{figure}

A sequence of current pulses through a tunnel barrier is shown schematically in Fig.\ \ref{fig_pulses}. Their characteristic width $\delta t= h/eV$ is much less than their mean spacing $\Delta t=e/I=h/2e\tau V$ if the tunnel probability $\tau\ll 1$. It is therefore possible in principle to detect the individual entangled electron-hole pairs and test for a violation of the Bell inequality by correlating their spins. In a low-frequency noise measurement a large number of current pulses is detected and the spin of each electron is correlated with the spin of every hole, regardless of whether the two particles were produced by the same tunnel event or not. Electrons and holes from different tunnel events have uncorrelated spins, so they do not contribute to the correlator. The low-frequency measurement is therefore equivalent to a time-resolved measurement. 

As formulated in Sec.\ \ref{Bell_ineq}, the CHSH inequality contains the two-particle correlator $C_{\bm{ab}}$ of spin $A$ measured along unit vector $\bm{a}$ and spin $B$ measured along unit vector $\bm{b}$. In a conductor one would measure correlators of currents rather than of individual particles. Chtchelkatchev, Blatter, Lesovik, and Martin \cite{Cht02} have reformulated the Bell inequality in these terms.

The correlator $K_{ij}=\langle N_{A,i}N_{B,j}\rangle$ of the number of electrons (holes) detected in a time $t_{\rm det}$ with spin $i$ ($j$) at detector $A$ ($B$) is obtained by integrating the current correlator,
\begin{eqnarray} 
K_{ij}&=&\int_{0}^{t_{\rm det}}dt\int_{0}^{t_{\rm det}}dt'\,\langle I_{A,i}(t)I_{B,j}(t')\rangle\nonumber\\
&=&t_{\rm det}^{2}\langle I_{A,i}\rangle\langle I_{B,j}\rangle+\int_{-\infty}^{\infty}d\omega\,\frac{2\sin^{2}(\omega t_{\rm det}/2)}{\pi\omega^{2}}C_{ij}(\omega).\nonumber\\
&&\label{Bellcorrelator}
\end{eqnarray}
Here $C_{ij}(\omega)$ is the frequency dependent correlator of current fluctuations \cite{Bla00},
\begin{eqnarray}
C_{ij}(\omega)&=&\int_{-\infty}^{\infty}dt\,e^{i\omega t}\langle\delta I_{A,i}(t)\delta I_{B,j}(0)\rangle\nonumber\\
&=&C_{ij}(0)\times\left\{\begin{array}{cl}
1-|\hbar\omega/eV|&\;\;{\rm if}\;\;|\hbar\omega/eV|<1,\\
0&\;\;{\rm if}\;\;|\hbar\omega/eV|>1.
\end{array}\right.\nonumber\\
&&\label{Cijdef}
\end{eqnarray}
We have assumed that $V$ is bigger than temperature but still small enough that the energy dependence of the scattering matrix may be neglected in the range $(E_{F},E_{F}+eV)$. The dependence of $K_{ij}$ on the unit vectors $\bm{a},\bm{b}$ (along which the spin is measured) is implicit. The subscripts $i,j\in\{\uparrow,\downarrow\}\equiv\{1,2\}$ may equally well refer to two orbital degrees of freedom, as explained in Sec.\ \ref{spinvsorbit}.

The correlator $C_{\bm{ab}}$ in the CHSH inequality (\ref{CHSHdef}) is determined by
\begin{eqnarray}
C_{\bm{ab}}&=&\frac{\langle(N_{A,1}-N_{A,2})(N_{B,1}-N_{B,2})\rangle}{\langle(N_{A,1}+N_{A,2})(N_{B,1}+N_{B,2})\rangle}\nonumber\\
&=&\frac{K_{11}+K_{22}-K_{12}-K_{21}}{K_{11}+K_{22}+K_{12}+K_{21}},\label{CabKrelation}
\end{eqnarray}
where the detection time $t_{\rm det}$ should be in the range $\delta t\ll t_{\rm det}\ll\Delta t$ to ensure coincidence detection. The average $\langle\cdots\rangle$ is over many measurements, of which a fraction $\nu=\langle(N_{A,1}+N_{A,2})(N_{B,1}+N_{B,2})\rangle$ is successful in detecting one particle at each detector $A$ and $B$. This non-unit detection efficiency appears in the denominator of Eq.\ (\ref{CabKrelation}), so that the resulting correlator counts only the successful measurements [in accordance with Eq.\ (\ref{CHSHnu})].

Because, on the one hand, $t_{\rm det}\gg\delta t=eV/h$ is sufficiently large one may approximate $2\sin^{2}(\omega t_{\rm det}/2)/\pi\omega^{2}\rightarrow t_{\rm det}\delta(\omega)$ in Eq.\ (\ref{Bellcorrelator}) by its long-time limit, hence
\begin{equation}
K_{ij}\rightarrow t_{\rm det}^{2}\langle I_{A,i}\rangle\langle I_{B,j}\rangle+t_{\rm det}C_{ij}(0).\label{KlongtimeC}
\end{equation}
On the other hand, because $t_{\rm det}\ll\Delta t=h/2e\tau V$ is sufficiently small one may neglect the first term on the right-hand-side of Eq.\ (\ref{KlongtimeC}) relative to the second term: $t_{\rm det}\langle I\rangle^{2}/C\simeq t_{\rm det}\langle I\rangle/e\ll 1$. The correlator $C_{\bm{ab}}$ in the CHSH inequality is then given entirely in terms of low-frequency current correlators,
\begin{equation}
C_{\bm{ab}}=\frac{C_{11}(0)+C_{22}(0)-C_{12}(0)-C_{21}(0)}{C_{11}(0)+C_{22}(0)+C_{12}(0)+C_{21}(0)}.\label{CabC0relation}
\end{equation}

To verify this line of argument, one can calculate the concurrence of the electron-hole pairs that follows from the violation of the CHSH inequality using the correlator (\ref{CabC0relation}) and compare with the expected result (\ref{Cdef4}). This calculation \cite{Bee03} is outlined in Appendix \ref{app_ctunneling}.

\subsection{Beyond the tunneling regime}
\label{detection_notunneling}

If the transmission probability $\tau$ is not $\ll 1$, then there is no separation of the time scales $\delta t$, $\Delta t$ and the argument from the previous sub-section breaks down. It is still possible to determine the degree of entanglement using low-frequency measurements, but a modified expression for the spin-spin correlator $C_{\bm ab}$ is needed \cite{Bee04a,Sam04}.

For short detection times $t_{\rm det}\ll h/eV$, shorter than both $\delta t$ and $\Delta t$, one may take the limit
\begin{widetext}
\begin{equation}
\lim_{t_{\rm det}\rightarrow 0}t_{\rm det}^{-2}\int_{-\infty}^{\infty}d\omega\,\frac{2\sin^{2}(\omega t_{\rm det}/2)}{\pi\omega^{2}}C_{ij}(\omega)=\int_{-\infty}^{\infty}\frac{d\omega}{2\pi}C_{ij}(\omega)=\frac{eV}{h}C_{ij}(0),\label{Cijzero}
\end{equation}
cf.\ Eq.\ (\ref{Cijdef}). Instead of Eq.\ (\ref{KlongtimeC}) we now have
\begin{equation}
K_{ij}\rightarrow t_{\rm det}^{2}[\langle I_{A,i}\rangle\langle I_{B,j}\rangle+(eV/h)C_{ij}(0)].\label{KshorttimeC}
\end{equation}
Substitution into Eq.\ (\ref{CabKrelation}) gives the spin-spin correlator
\begin{equation}
C_{\bm{ab}}=\frac{(h/eV)\langle I_{A,1}-I_{A,2}\rangle\langle I_{B,1}-I_{B,2}\rangle+C_{11}(0)+C_{22}(0)-C_{12}(0)-C_{21}(0)}{(h/eV)\langle I_{A,1}+I_{A,2}\rangle\langle I_{B,1}+I_{B,2}\rangle+C_{11}(0)+C_{22}(0)+C_{12}(0)+C_{21}(0)}.\label{CabC0relation2}
\end{equation}
\end{widetext}

The correlator (\ref{CabC0relation2}) differs from the expression (\ref{CabC0relation}) in the tunneling regime by the extra terms containing the product of average currents. It still contains only low-frequency quantities, so it can be evaluated without requiring time-resolved detection. In Appendix \ref{app_cnotunneling} we show that the correlator (\ref{CabC0relation2}) produces again the expected result (\ref{Cdef4}) for the concurrence, now without the requirement $\tau_{1},\tau_{2}\ll 1$.

\subsection{Full counting statistics}
\label{fullcounting}

The relation between entanglement and current fluctuations has been extended to higher than second order correlators by Faoro, Taddei, and Fazio \cite{Fao04} and by Di Lorenzo and Nazarov \cite{Lor05}. This is the problem of full counting statistics. A complication in the analysis is that electrons do not traverse a quantum conductor as separate and independent particles unless the transmission probability is vanishingly small. There is no way to determine {\em a priori\/} if and how the electron flow can be partitioned into elementary events involving one, two, or larger number of electrons. In Ref.\ \cite{Lor05}, using the spin entangler of Fig.\ \ref{fig_geometries_spin}c, it was shown that the elementary events of electron transfer which contribute to the full counting statistics can be viewed as single-electron and two-electron events. Two electrons are always transferred as the spin singlet (\ref{Belldef}). The fraction of electrons transferred as singlets depends on the transmission probability. While all electrons come in singlets for a perfectly transmitting channel, no singlets are transferred in the limit of vanishing transmission probability.

\section{Loss of entanglement by dephasing}
\label{dephasing}

In Sec.\ \ref{free_fermions} we discussed how thermal fluctuations degrade the entanglement of the electron-hole pair. The pure state at zero temperature becomes a mixed state as a result of the thermal fluctuations. Beyond a critical temperature $T_{c}$ of order $eV/k_{B}$ the entanglement production at the tunnel barrier vanishes identically (cf.\ Fig.\ \ref{fig_Epartplot}).

If $T\ll eV/k_{B}$ thermal fluctuations are irrelevant, but the electron-hole state may still have lost its purity (= have dephased or decohered) by other mechanisms. The two types of entanglement discussed in Sec.\ \ref{spinvsorbit}, spin or orbital, dephase for different reasons. For spin entanglement, the spin-orbit interaction and the hyperfine interaction with nuclear spins are effective mechanisms of dephasing. For orbital entanglement, the electromagnetic fluctuations caused by other charges lead to dephasing. In this section we concentrate on the latter case, specifically in the edge channel geometry of Fig.\ \ref{fig_geometries_orbit}b.

The dephasing by electromagnetic fluctuations of entangled electron-hole pairs in spatially separated edge channels has been studied in Refs.\ \cite{Bee03,Sam03,Vel03,Sam03t}. Here we follow Ref.\ \cite{Vel03}. We model the effect of the electromagnetic fluctuations phenomenologically by introducing phase factors in the scattering matrix and subsequently averaging over these phases. We refer to Ref.\ \cite{Pra05} for an alternative phenomenological approach (using a dephasing voltage probe). A more microscopic treatment of the effect of dephasing on electron-hole entanglement (for example along the lines set out by Marquardt and Bruder \cite{Mar04}) has not yet been attempted.

Introducing random phase shifts $\phi_{i}$ ($\psi_{i}$) accumulated in channel $i$ at the left (right) of the tunnel barrier, the reflection and transmission matrices transform as
\begin{equation}
r\rightarrow\begin{pmatrix}
e^{i\phi_{1}} & 0 \\
0 & e^{i\phi_{2}} 
\end{pmatrix}
r_{0},\;\;t\rightarrow\begin{pmatrix} 
e^{i\psi_{1}} & 0 \\
0 & e^{i\psi_{2}} 
\end{pmatrix}t_{0}.\label{rtepsilon}
\end{equation}
By averaging over the phase shifts, with distribution $P(\phi_{1},\phi_{2},\psi_{1},\psi_{2})$, the pure state $|\Psi_{\rm out}\rangle$ [given by Eq.\ (\ref{Psioutdef})] is converted into the mixed density matrix
\begin{equation}
\rho_{\rm out}=\int d\phi_{1}d\phi_{2}d\psi_{1}d\psi_{2}\,P(\phi_{1},\phi_{2},\psi_{1},\psi_{2})|\Psi_{\rm out}\rangle\langle\Psi_{\rm out}|.\label{rhooutmixed}
\end{equation}
To simplify the average, the phase shifts at the left and the right of the tunnel barrier are assumed to be independent with identical distributions, so
$P(\phi_{1},\phi_{2},\psi_{1},\psi_{2})=P(\phi_{1},\phi_{2})P(\psi_{1},\psi_{2})$. The complex dephasing parameter $\eta$ is defined by
\begin{equation}
\eta=\int d\phi d\phi'\, P(\phi,\phi')e^{i\phi-i\phi'}.\label{etadef}
\end{equation}

Analytical progress can be made in the case that the two transmission eigenvalues (eigenvalues of $tt^{\dagger}$) are identical: $\tau_{1}=\tau_{2}\equiv \tau$. In the absence of dephasing the electron and hole then form a maximally entangled Bell pair, with concurrence ${\cal C}=1$. The matrix product $r_{0}\sigma_{y}t_{0}^{T}$, which appears in the expression (\ref{Psioutdef}) for $|\Psi_{\rm out}\rangle$, is parameterized by
\begin{equation}
r_{0}\sigma_{y}t_{0}^{T}=\sqrt{\tau(1-\tau)}\begin{pmatrix}
\cos\xi&\sin\xi\\
-\sin\xi&\cos\xi
\end{pmatrix}.\label{Omegadef}
\end{equation}
The angle $\xi$ governs the extent to which the tunnel barrier mixes the edge channels (no mixing for $\xi=0,\pi/2$, complete mixing for $\xi=\pi/4$). 

Just as in the absence of dephasing (cf.\ App.\ \ref{app_spindep}) the entanglement production is determined entirely by the projection of the density matrix $\rho_{\rm out}$ on a single particle left and right of the tunnel barrier. The projected density matrix $\rho_{11}$ has elements
\begin{equation}
\rho_{11}=\frac{1}{4}\begin{pmatrix}
2\cos^{2}\xi&\eta\sin 2\xi& -\eta^{\ast}
\sin 2\xi& 2|\eta|^{2}\cos^{2}\xi\\
\eta^{\ast}\sin 2\xi&2\sin^{2}\xi& -2\eta^{\ast\,2}\sin^{2}\xi& \eta^{\ast}\sin 2\xi\\
-\eta\sin 2\xi& -2\eta^{2} \sin^{2}\xi&2\sin^{2}\xi& -\eta\sin 2\xi\\
2|\eta|^{2}\cos^{2}\xi& \eta\sin 2\xi&  -\eta^{\ast}\sin 2\xi&2\cos^{2}\xi
\end{pmatrix}.\label{rhoresult}
\end{equation}
The weight $w_{11}=2\tau(1-\tau)$ of the projection is independent of dephasing.

From the mixed density matrix (\ref{rhoresult}) we calculate the concurrence  using Wootters formula (\ref{Cdef2}), with the result
\begin{eqnarray}
{\cal C}&=&{\rm max}\bigl\{0,-\tfrac{1}{2}(1-|\eta|^{2})\nonumber\\
&&\mbox{}+\tfrac{1}{4}\sqrt{16|\eta|^{2}+2(1-|\eta|^{2})^{2}(1+\cos 4\xi)} \bigr\}.
\label{cfinal}
\end{eqnarray}
Notice that ${\cal C}=|\eta|^{2}$ for $\xi=0$. The entanglement production rate follows from 
\begin{equation}
{\cal E}_{\rm part}/t_{\rm det}=\frac{eV}{h}2\tau(1-\tau){\cal F}\left(\tfrac{1}{2}+\tfrac{1}{2}\sqrt{1-{\cal C}^{2}}\right),\label{Epartdephasing}
\end{equation}
with the function ${\cal F}$ defined in Eq.\ (\ref{calFdef}).

\begin{figure}
\begin{center}
\includegraphics[width=0.8\linewidth]{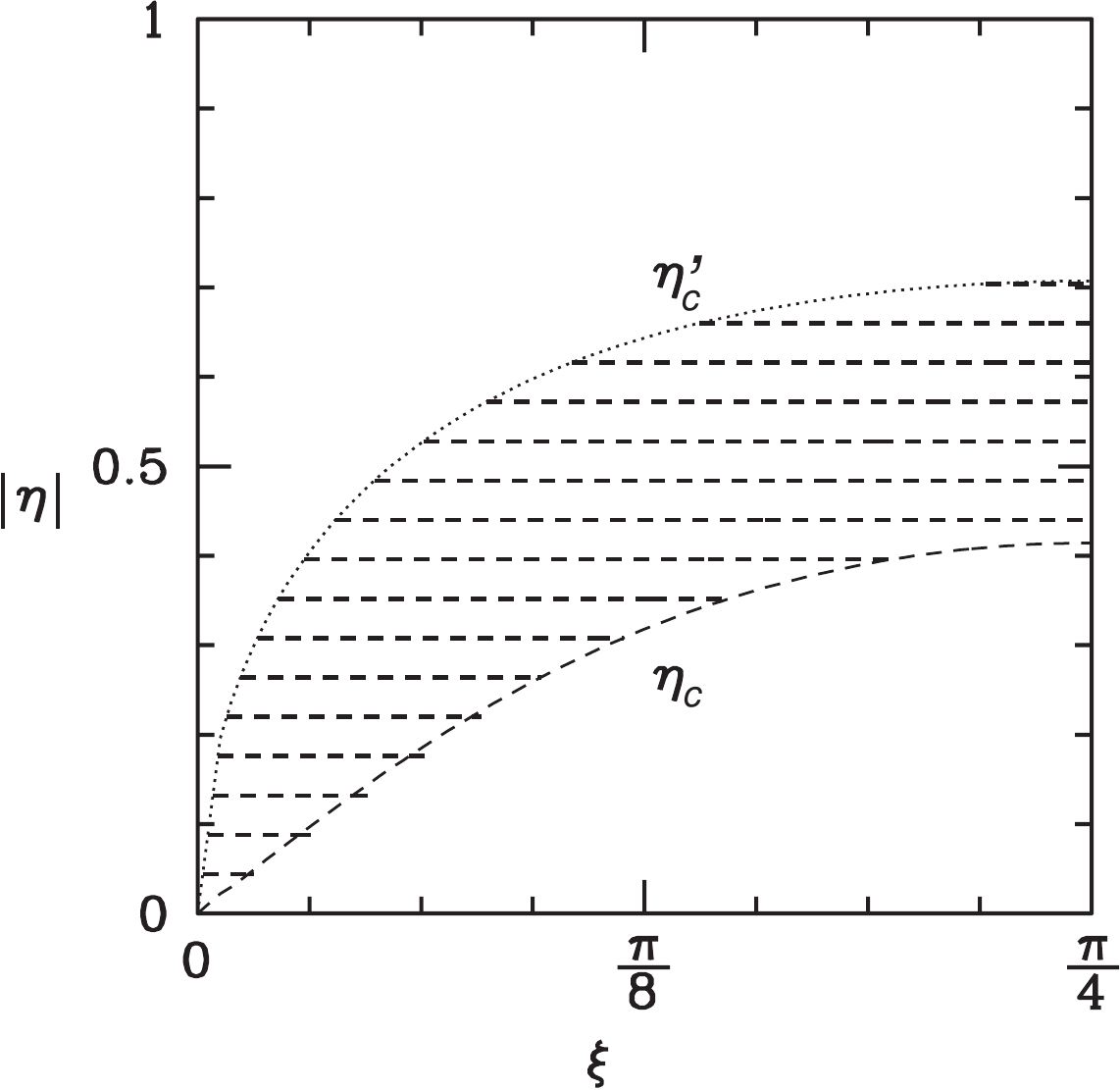}
\end{center}
\caption{The Bell inequality is violated for dephasing parameters $|\eta| > \eta'_{c}$, while entanglement is preserved for $|\eta| > \eta_{c}$. The dashed and dotted curves are computed from Eqs.\ (\ref{etacresult}) and (\ref{etacprime}), respectively. The shaded region between the two curves indicates dephasing and mixing parameters for which there is entanglement without violation of the Bell inequality. Figure from Ref.\ \cite{Vel03}. \label{deph}
}
\end{figure}

Instead of a critical temperature we now have a critical dephasing parameter $\eta_{c}$, such that the entanglement production vanishes identically for $|\eta|\le \eta_{c}$. (Smaller $|\eta|$ means stronger dephasing.) From Eq.\ (\ref{cfinal}) one finds the expression
\begin{equation}
\eta_{c}=\sqrt{\frac{5-\cos 4\xi -2\sqrt{6-2\cos 4\xi}}{1-\cos 4\xi}},\label{etacresult}
\end{equation}
plotted in Fig.\ \ref{deph} (dashed curve). Unlike the critical temperature (\ref{Tceq}), the critical dephasing parameter is independent of the tunnel probability $\tau$. It does depend sensitively on the mixing parameter $\xi$, increasing from $0$ to $\sqrt{2}-1$ as $\xi$ increases from $0$ to $\pi/4$.

\begin{figure}
\includegraphics[width=0.8\linewidth]{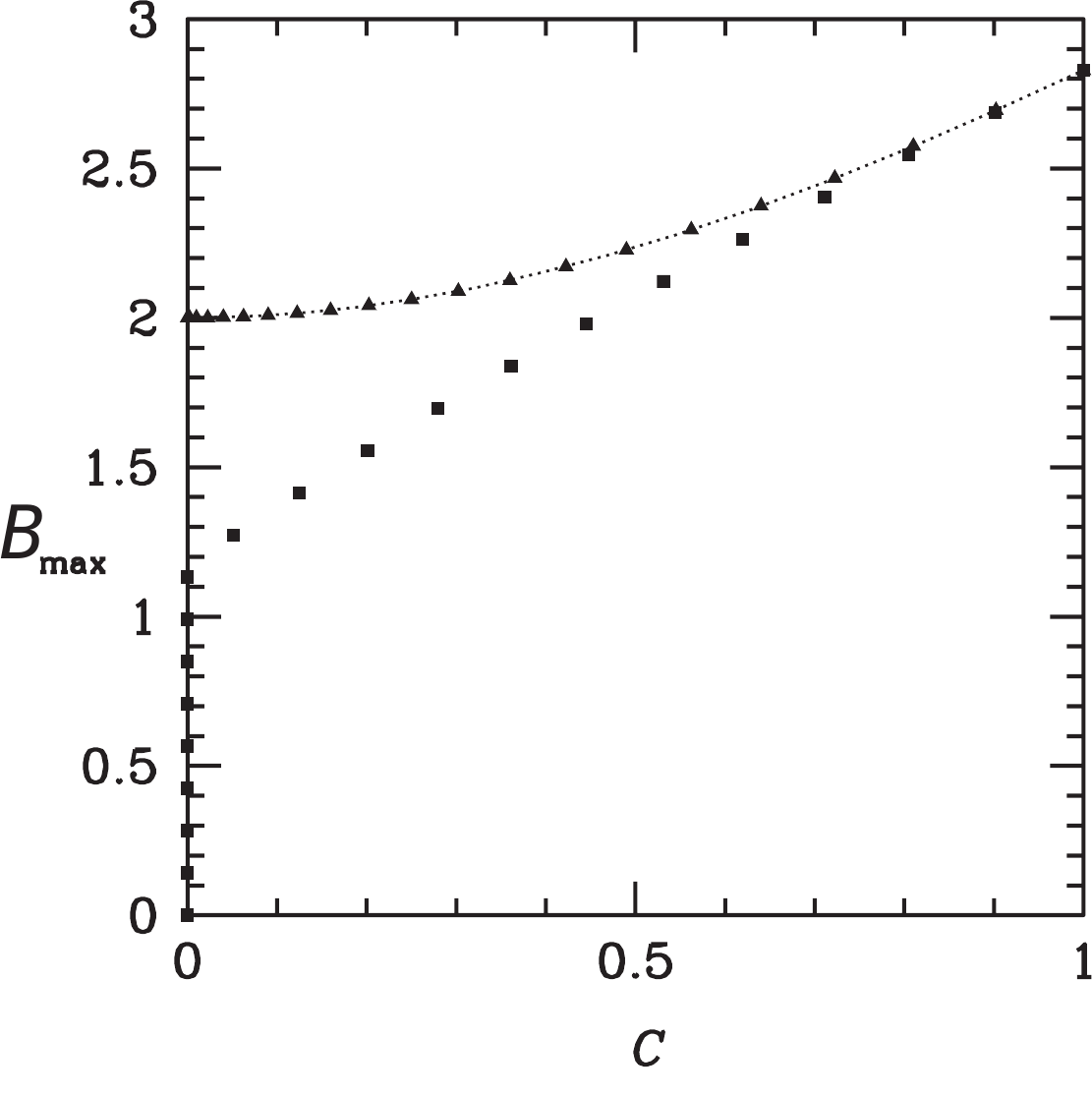}
\caption{
Relation between the maximal violation ${\cal B}_{\rm max}$ of the Bell inequality and the concurrence ${\cal C}$, calculated from Eqs.\ (\ref{cfinal}) and (\ref{efinal}) for mixing parameters $\xi=0$ (triangles, no mixing) and $\xi=\pi/4$ (squares, complete mixing). The dephasing parameter $|\eta|$ decreases from 1 (upper right corner, no dephasing) to 0 (lower left, complete dephasing) with steps of 0.05. The dotted line is the relation between ${\cal B}_{\rm max}$ and ${\cal C}$ for a pure state, which is also the largest possible value of ${\cal B}_{\rm max}$ for given ${\cal C}$. Figure from Ref.\ \cite{Vel03}. \label{cbell}
}
\end{figure}

As discussed in Sec.\ \ref{meas_mixed}, for a mixed state there is no one-to-one relation between the Bell parameter ${\cal B}_{\rm max}$ and the concurrence ${\cal C}$. Application of the general formula (\ref{Bmixed}) to the density matrix (\ref{rhoresult}) gives
\begin{equation}
{\cal B}_{\rm max}=\sqrt{2(1+|\eta|^{2})^{2}+2(1-|\eta|^{2})^{2}\cos 4\xi}.
\label{efinal}
\end{equation}
The Bell inequality ${\cal B}_{\rm max}\leq 2$ is violated for $|\eta|\geq\eta'_{c}$, with
\begin{equation}
\eta'_{c}=\sqrt{\frac{-1+\cos 4\xi +\sqrt{2-2\cos 4\xi}}{1+\cos 4\xi}} \label{etacprime}
\end{equation}
increasing from $0$ to $\sqrt{2}$ as $\xi$ increases from $0$ to $\pi/4$ (dotted curve in Fig.\ \ref{deph}).

In Fig.\ \ref{cbell} we compare ${\cal B}_{\rm max}$ and ${\cal C}$ for $\xi=0$ (no mixing) and $\xi=\pi/4$ (complete mixing). For $\xi=0$ the same relation (\ref{BmaxCrelation}) between ${\cal B}_{\rm max}$ and ${\cal C}$ holds as for pure states (dotted curve). Violation of the Bell-CHSH inequality (\ref{CHSHdef}) is then equivalent to entanglement. For $\xi\neq 0$ there exist entangled states (${\cal C}>0$) without violation of the Bell inequality (${\cal B}_{\rm max} \le 2$). This is indicated by the shaded region in Fig.\ \ref{deph}. For $\xi\neq 0$ violation of the Bell inequality is thus a sufficient but not a necessary condition for entanglement.

\section{Quantum entanglement pump}
\label{epump}

\begin{figure}
\includegraphics[width=0.8\linewidth]{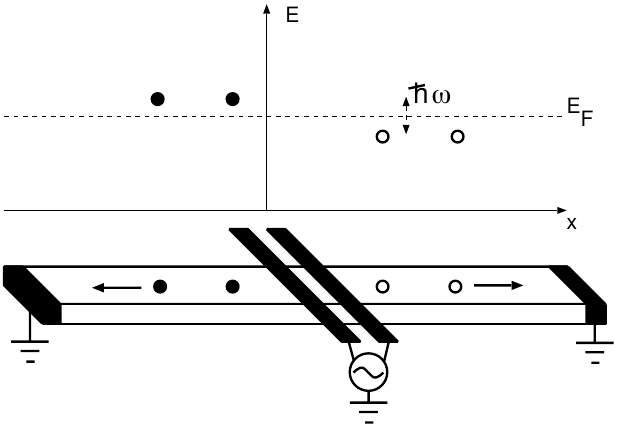}
\caption{
Production of entangled electron-hole pairs in a narrow ballistic conductor by a quantum pump. The left and right ends of the conductor are at the same potential, while the potential on the gate electrodes at the center is periodically modulated. Such a device produces spatially separated electron-hole pairs (black and white circles), differing in energy by a multiple of the pump frequency $\omega$. For spin-independent scattering the electron ($e$) and hole ($h$) produced during a given cycle have the same spin $\uparrow,\downarrow$, so that their wave function is that of a Bell pair, $\propto|\!\uparrow_{e}\uparrow_{h}\rangle+|\!\downarrow_{e}\downarrow_{h}\rangle$. The optimal quantum entanglement pump produces, on average, $1$ Bell pair every $2$ cycles. Figure from Ref.\ \cite{Bee05}.
}
\label{fig_epump}
\end{figure}

The quantum electron pump is a device that transfers electrons phase coherently between two reservoirs at the same voltage, by means of a slowly oscillating voltage on a gate electrode \cite{Swi99}. Special pump cycles exist that transfer the charge in a quantized fashion, one $e$ per cycle. Samuelsson and B\"{u}ttiker \cite{Sam05} asked the question whether a similar device could be used to create entangled electron-hole pairs in a controlled manner, clocked by the gate voltage, rather than stochastically as in the original proposal \cite{Bee03}. (See Fig.\ \ref{fig_epump} for a schematic illustration.) Such a quantum entanglement pump could be a building block of quantum computing designs using ballistic flying qubits in nanowires or in quantum Hall edge channels \cite{Ber00,Ion01,Bar04}.

We already noticed in Sec.\  \ref{free_fermions} [cf.\ Eq.\ (\ref{shotnoisepower})] that the rate of entanglement production is closely related to the charge noise, to the extent that a noiseless scatterer produces no entanglement. By maximizing the charge noise with spin-independent scattering we calculated \cite{Bee05} that a pump can at most produce, on average, 1 Bell pair every 2 cycles. (A similar conclusion was reached in Ref.\ \cite{Leb05b}.) A deterministic spin-entangler \cite{DiV04}, being the analogue of a quantized charge pump, would have an entanglement production of $1$ Bell pair per cycle, so the optimal entanglement pump has one half the efficiency of a deterministic entangler.

\section{Teleportation by electron-hole annihilation}
\label{teleport}

\begin{figure}
\includegraphics[width=0.8\linewidth]{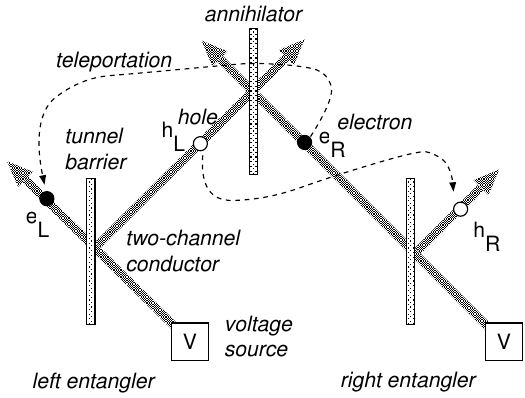}
\caption{
Schematic description of teleportation by electron-hole annihilation. A voltage $V$ applied over a tunnel barrier produces pairs of entangled electron-hole pairs in the Fermi sea. One such pair $(e_{L},h_{L})$ is shown at the left. For a simplified description we assume spin entanglement in the state $(|\!\uparrow\uparrow\rangle+|\!\downarrow\downarrow\rangle)/\sqrt{2}$, where the first arrow refers to the electron spin and the second arrow to the hole spin. A second electron $e_{R}$ is in an unknown state $\alpha|\!\uparrow\rangle+\beta|\!\downarrow\rangle$. The electron $e_{R}$ can annihilate with the hole $h_{L}$ by tunneling through the barrier at the center. If it happens, the state of $e_{L}$ collapses to the state of $e_{R}$. (Notice that $|\!\uparrow\rangle$ annihilates with $|\!\uparrow\rangle$ and $|\!\downarrow\rangle$ annihilates with $|\!\downarrow\rangle$, so $e_{L}$ inherits the coefficients $\alpha$ and $\beta$ of $e_{R}$ after its annihilation.) The diagram shows a second entangler at the right, to perform two-way teleportation (from $e_{R}$ to $e_{L}$ and from $h_{L}$ to $h_{R}$).  This leads to entanglement swapping: $e_{L}$ and $h_{R}$ become entangled after the annihilation of $h_{L}$ and $e_{R}$. Figure from Ref.\ \cite{Bee04b}.
\label{teleport_schematic}
}
\end{figure}

\begin{figure*}
\includegraphics[width=0.8\linewidth]{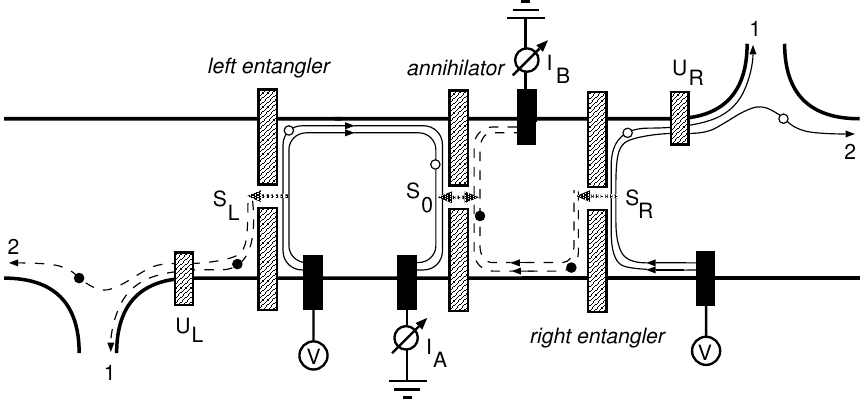}
\caption{
Proposed realization of the teleportation scheme of Fig.\ \ref{teleport_schematic}, using edge channels in the quantum Hall effect. The thick black lines indicate the boundaries of a two-dimensional electron gas, connected by Ohmic contacts (black rectangles) to a voltage source $V$ or to ground. A strong perpendicular magnetic field ensures that the transport in an energy range $eV$ above the Fermi level takes place in two edge channels, extended along a pair of equipotentials (thin solid and dashed lines, with arrows that give the direction of propagation). These edge channels realize the two-channel conductors of Fig.\ \ref{teleport_schematic}, with the Landau level index $n=1,2$ playing the role of the spin index $\uparrow,\downarrow$. Solid lines signify predominantly filled edge channels with hole excitations (open circles), while dashed lines signify predominantly empty edge channels with electron excitations (black dots). The beam splitters of Fig.\ \ref{teleport_schematic} are formed by split gate electrodes (shaded rectangles), through which the edge channels may tunnel (dashed arrows, scattering matrices $S_{L},S_{R},S_{0}$). The annihilation of the electron-hole excitation at the central beam splitter is detected through the currents $I_{A}$ and $I_{B}$. Entanglement swapping resulting from two-way teleportation is detected by the violation of a Bell inequality. This requires two gate electrodes to locally mix the edge channels (scattering matrices $U_{L}$, $U_{R}$) and two pair of contacts $1,2$ to separately measure the current in each transmitted and reflected edge channel. Notice that there are no phase-coherent paths connecting the left and right ends of the conductor (because of the intervening dephasing contacts A and B), so a demonstration of entanglement between the two ends is indeed a demonstration of teleportation. Figure from Ref.\ \cite{Bee04b}.
\label{teleport_QHE}
}
\end{figure*}

Teleportation is the disembodied transport of a quantum mechanical state between two locations that are only coupled by classical (incoherent) communication \cite{Ben93}. What is required is that the two locations share a previously entangled state. Teleportation has the remarkable feature that the teleported state need not be known. It could even be undefined as a single-particle state, which happens if the teleported particle is entangled with another particle that stays behind. Teleportation then leads to ``entanglement swapping'' \cite{Yur92,Zuk93}: Pre-existing entanglement is exchanged for entanglement between two parties that have never met.

Experiments with photons \cite{exp} have demonstrated that teleportation can be realized in practice. Only {\em linear\/} optical elements are needed \cite{Vai99,Lut99}, if one is satisfied with a success probability $<1$. Such non-deterministic teleportation plays an essential role in proposals for a quantum computer based entirely on linear optics \cite{Kni01}. A central requirement for nontrivial logical operations is that the linear elements (beam splitters, phase shifters) are supplemented by single-photon sources and single-photon detectors, which effectively introduce nonlinearities. 

Teleportation of electrons has not yet been realized. The analogue of teleportation by linear optics would be teleportation of free electrons, that is to say, teleportation using only single-particle Hamiltonians. A direct translation of existing linear optics protocols would require single-electron sources and single-electron detectors \cite{Sau03}. Such devices exist, but not for free electrons --- they are all based on the Coulomb interaction in quantum dots.

In this section we review, following Ref.\ \cite{Bee04b}, an alternative single-particle teleportation mechanism using entangled electron-hole pairs.
The key observation is that the annihilation of an electron-hole pair in the Fermi sea teleports these quasiparticles to a distant location, if entanglement was established beforehand. This two-way teleportation scheme is explained in Fig.\ \ref{teleport_schematic}.

The analysis is simplest for the entangled state
\[
(|\!\uparrow\rangle_{e}|\!\uparrow\rangle_{h}+ |\!\downarrow\rangle_{e}|\!\downarrow\rangle_{h})/\sqrt{2}.
\]
The subscripts $e$ and $h$ refer, respectively, to electron and hole at two distant locations. The particle to be teleported is another electron, in the state $\alpha|\!\uparrow\rangle_{e'}+\beta|\!\downarrow\rangle_{e'}$ (with $|\alpha|^{2}+|\beta|^{2}=1$). The second electron $e'$ may tunnel into the empty state representing the hole $h$, but only if the spins match. If $\tau$ denotes the tunneling amplitude, then this happens with probability $\frac{1}{2}|\alpha|^{2}|\tau|^{2}+\frac{1}{2}|\beta|^{2}|\tau|^{2}=\frac{1}{2}|\tau|^{2}\ll 1$. The resulting annihilation of the two quasiparticle excitations collapses the combined state
\[
(\alpha|\!\uparrow\rangle_{e'}+\beta|\!\downarrow\rangle_{e'})
(|\!\uparrow\rangle_{e}|\!\uparrow\rangle_{h}+ |\!\downarrow\rangle_{e}|\!\downarrow\rangle_{h})/\sqrt{2}
\]
to the state $\alpha|\!\uparrow\rangle_{e}+\beta|\!\downarrow\rangle_{e}$, so the state of the second electron $e'$ is teleported to the first electron $e$ at a distant location.

The usual limitations \cite{Ben93} of teleportation apply. Since tunneling is an unpredictable stochastic event, it has to be detected and communicated (by classical means) to the distant location. There is therefore no instantaneous transfer of information. And since the electron has to be annihilated in order to be teleported, its state can not be copied. Teleportation by electron-hole annihilation thus presents a rather dramatic demonstration of the no-cloning theorem of quantum mechanics \cite{Woo82,Die82}.

A major obstacle to teleportation in the solid state is the requirement of fast time-resolved detection. To circumvent this difficulty Ref.\ \cite{Bee04b} identified a low-frequency noise correlator that demonstrates the entanglement swapping resulting from two-way teleportation. Two-way teleportation means that upon annihilation the electron and hole are teleported to opposite ends of the system. The noise correlator measures the degree of entanglement at the two ends. This demonstrates teleportation if the two ends are not connected by any phase-coherent path.

In Fig.\  \ref{teleport_QHE} a particular implementation is illustrated using edge channels in the quantum Hall effect regime. To detect the entanglement swapping one needs to measure a fourth order cumulant of fluctuations of tunneling current. Typically, only the second order cumulant is measured in noise experiments, but higher order cumulants are now becoming accessible as well \cite{Reu03,Bom05}.

\section{Three-qubit entanglement}
\label{three-qubit}

\begin{figure}
\includegraphics[width=0.8\linewidth]{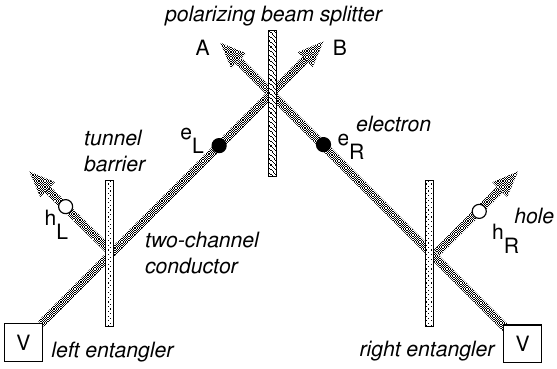}
\caption{
Schematic description of the creation of three-qubit entanglement out of two
entangled electron-hole pairs in the Fermi sea. The left and right entangler
consist of a tunnel barrier over which a voltage $V$ is applied. For a
simplified description we assume spin entanglement in the state
$(|\!\uparrow_{h}\uparrow_{e}\rangle+|\!\downarrow_{h}\downarrow_{e}\rangle)
/\sqrt{2}$, where
the subscripts $e,h$ refer to electron and hole spin. The two electrons
meet at a polarizing beam splitter, which fully transmits the up-spin and fully
reflects the down-spin. If the outgoing ports A, B contain one electron each,
then they must both have the same spin. The corresponding outgoing state has
the form
$(|\!\uparrow_{h}\uparrow_{h}\uparrow_{e}\uparrow_{e}\rangle+
|\!\downarrow_{h}\downarrow_{h}\downarrow_{e}\downarrow_{e}\rangle)/\sqrt{2}$. 
Since the two electrons at A,B are constrained to have the same spin, 
this four-particle GHZ state represents three independent logical qubits. Figure from Ref.\ \cite{Bee04c}.
\label{GHZ_schematic}
}
\end{figure}

\begin{figure*}
\includegraphics[width=0.8\linewidth]{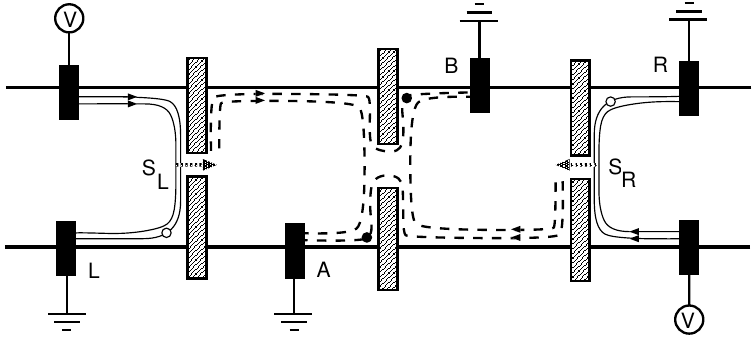}
\caption{
Proposed realization of the three-qubit entangler, using edge channels in the
quantum Hall effect. The left and right point contacts (scattering matrices
$S_{L}$, $S_{R}$) each produce entangled electron-hole pairs in the Fermi sea.
They partially transmit and reflect both edge channels, analogously to beam
splitters in optics. The central point contact is the analogue of a polarizing
beam splitter: It fully transmits the inner edge channel and fully reflects the
outer one. Three-qubit entanglement results if there is one excitation at each
of the four edges $L,R,A,B$. The two electron excitations at $A$ and $B$ then
have the same channel index, so they constitute a {\em single\/} qubit. This
qubit forms a three-qubit entangled state with the two hole excitations at $L$
and $R$. Figure from Ref.\ \cite{Bee04c}.
\label{GHZ_QHE}
}
\end{figure*}

So far we have always considered bipartite entanglement. The interaction-free entanglement of three qubits (= tripartite entanglement) was investigated in Ref.\ \cite{Bee04c}.

The proposed three-qubit entangler is sketched schematically in Fig.\
\ref{GHZ_schematic}. As in the original three-photon entangler of Zeilinger
{\em et al.} \cite{Zei97}, three-qubit entanglement is created out of
two entangled electron-hole pairs. Fig.\ \ref{GHZ_QHE} shows a physical realization of Fig.\
\ref{GHZ_schematic}, using edge channels in the quantum Hall effect.

The irreducible tripartite entanglement is quantified by the tangle $\tau$ of
Coffman, Kundu, and Wootters \cite{Cof00}, which is the three-qubit analogue of
the concurrence. The tangle is unity for the maximally entangled
Greenberger-Horne-Zeilinger (GHZ) state $|\uparrow\uparrow\uparrow\rangle+|\downarrow\downarrow\downarrow\rangle$ and vanishes if one qubit is
disentangled from the other two. Unlike the concurrence (\ref{Cdef4}), the tangle
depends not only on the transmission eigenvalues of the point contact
entanglers, but also on the eigenvectors. 

As in the bipartite case, current correlators can be used to test for the maximal violation of a tripartite generalization of the Bell inequality. Two tripartite
inequalities are known, one due to Mermin \cite{Mer90} and the other to
Svetlichny \cite{Sve87}.
While there exists a one-to-one relation (\ref{BmaxCrelation}) between concurrence and Bell
inequality for any pure state of two qubits, no such relation exists for $\tau$. A numerical investigation \cite{Ema03} has found a
simple set of upper and lower bounds for $\tau$. Since these bounds become
tighter and tighter as the state approaches the GHZ state, they should be of
practical use.

The bipartite electron-hole entangler \cite{Bee03} discussed in previous sections is capable of producing the most general
two-qubit entangled pure state, by suitably choosing the scattering matrix of
the tunnel barrier. The tripartite entangler \cite{Bee04c} discussed here, in contrast, is limited to the
production of a three-parameter subset of the most
general five-parameter family of three-qubit entangled pure states
\cite{Aci00}. This subset is characterized by the property that tracing over
the third qubit results in a mixed two-qubit state which is not entangled. The
origin of this restriction is that the three-qubit state is constructed out of
two separate entangled electron-hole pairs.

The two-qubit entangler can produce maximally entangled Bell pairs as well as
partially entangled states, as quantified by the concurrence. Similarly, the
three-qubit entangler can produce maximally entangled GHZ states as well as
states that have a smaller degree of tripartite entanglement, as quantified by
the tangle. However, in the three-qubit case there is a second
class of states that are irreducibly entangled and can not be obtained from the
GHZ state by any local operation \cite{Dur00}. These socalled W-states are not
accessible by the method of Fig.\ \ref{GHZ_schematic}. It would be interesting to see if there exists an alternative
interaction-free method to extract the W-state out of the Fermi sea, or whether
this is impossible as a matter of principle.

\section{The experimental challenge}
\label{experiments}

Tunneling experiments in metals and semiconductors have been carried out in the laboratory since the late 1950's, following the pioneering work of Esaki and Giaever \cite{Nobel}. The prediction of Ref.\ \cite{Bee03} is that at low temperatures these experiments produce electron-hole pairs in the spin-entangled state (\ref{Belldef}). So we may well have been producing Bell pairs in the solid state even earlier than in optics! The reason that the Bell inequality has been violated in optics \cite{Asp81} but not yet in electronics seems, therefore, to be a difficulty of detection rather than production. It is indeed much easier to measure photon polarization than electron spin.

Some of the implementations discussed in Sec.\ \ref{spinvsorbit} do not require spin-resolved detection, but since the entanglement then involves orbital degrees of freedom it is much more sensitive to decoherence. Attempts to observe two-particle interference effects of charge currents (for example, in the Mach-Zehnder geometry of Ref.\ \cite{Ji03}) have so far been unsuccesful, presumably because the electron charge is too strongly coupled to the environment \cite{Chu05}. Average spin currents can now be reliably detected \cite{Jed02}. It seems likely that correlators of spin currents will become accessible experimentally in the near future, permitting a test for the predicted electron-hole entanglement in the Fermi sea. 

\acknowledgments
I gratefully acknowledge my collaborators in this research, C. Emary, M. Kindermann, M. Titov, B. Trauzettel, and J. L. van Velsen, as well as the support by the Dutch Science Foundation NWO/FOM. I have received helpful comments on a draft of this review by S. D. Bartlett, S. J. van Enk, Yu.\ V. Nazarov, and P. Samuelsson.

\appendix

\section{Entanglement production for spin-dependent scattering}
\label{app_spindep}
In this appendix we review the calculation leading to Eqs.\ (\ref{ECrelation3}) and (\ref{Cdef4}), following Ref.\ \cite{Bee03}.

At zero temperature the incident state has the form
\begin{equation}
|\Psi_{\rm in}\rangle=\prod_{E_{F}<E<E_{F}+eV}a_{L,1}^{\dagger}(E)a_{L,2}^{\dagger}(E)|0\rangle. \label{Psiindef}
\end{equation}
The fermion creation operator $a^{\dagger}_{L,n}(E)$ (with $n=1,2\equiv\uparrow$, $\downarrow$) excites the $n$-th channel incident from the left at energy $E$. Similarly, $a^{\dagger}_{R,n}(E)$ excites a channel incident from the right. The product over energies refers to a discretization with interval $\delta E=h/t_{\rm det}$. The total number of energy intervals that contribute is $eV/\delta E=eVt_{\rm det}/h$.

It is convenient to collect the creation operators in two vectors $\bm{a}^{\dagger}_{L}$, $\bm{a}^{\dagger}_{R}$ and to use a matrix notation,
\begin{equation}
|\Psi_{\rm in}\rangle=
\prod_{E}
\begin{pmatrix}
\bm{a}_{L}^{\dagger}\\ \bm{a}_{R}^{\dagger}
\end{pmatrix}\begin{pmatrix}
\frac{1}{2}\sigma_{y}&0\\0&0
\end{pmatrix}\begin{pmatrix}
\bm{a}_{L}^{\dagger}\\ \bm{a}_{R}^{\dagger}
\end{pmatrix}|0\rangle.
\label{Psiinmatrix}
\end{equation}
The input-output relation (\ref{bSa}) can then be written as
\begin{equation}
\begin{pmatrix}
\bm{b}_{L}\\ \bm{b}_{R}
\end{pmatrix}
=\begin{pmatrix}
r&t'\\t&r'
\end{pmatrix}\begin{pmatrix}
\bm{a}_{L}\\ \bm{a}_{R}
\end{pmatrix}.\label{inputoutput}
\end{equation}
The $4\times 4$ unitary scattering matrix $S$ has $2\times 2$ submatrices $r,r',t,t'$ that describe reflection and transmission of states incident from the left or from the right. The transmission eigenvalues $\tau_{1},\tau_{2}\in[0,1]$ are the eigenvalues of $t^{\dagger}t=\openone-r^{\dagger}r$ (which are the same as the eigenvalues of $t'^{\dagger}t'=\openone-r'^{\dagger}r'$ because of unitarity). Substitution of Eq.\ (\ref{inputoutput}) into Eq.\ (\ref{Psiinmatrix}) gives the outgoing state
\begin{eqnarray}
|\Psi_{\rm out}\rangle&=&\prod_{E}\bigl(\bm{b}_{L}^{\dagger}
r\sigma_{y} t^{T}\bm{b}_{R}^{\dagger}
+[r\sigma_{y} r^{T}]_{12}^{\vphantom\dagger} b_{L,1}^{\dagger}b_{L,2}^{\dagger}\nonumber\\
&&\mbox{}+[t\sigma_{y} t^{T}]_{12}^{\vphantom\dagger} b_{R,1}^{\dagger}b_{R,2}^{\dagger}\bigr)|0\rangle. \label{Psioutdef}
\end{eqnarray}
The superscript $T$ indicates the transpose of a matrix.

As discussed in Sec.\ \ref{c_cons}, to calculate the accessible entanglement we need to account for the constraint of local particle number conservation in each energy interval $\delta E$.\footnote{One could relax this constraint, requiring only that the total particle number, summed over all energies, is conserved at the left and at the right of the tunnel barrier. The constraint used here, of particle conservation in each energy interval $\delta E$, is appropriate for an observer who uses only elastic scattering to detect the entanglement production.} 
The three terms at the right-hand-side of Eq.\ (\ref{Psioutdef}) represent different local particle numbers $p$ and $q$ at the left and right of the tunnel barrier, so the entanglement entropy of each term may be added to obtain the total entanglement entropy ${\cal E}_{\rm part}$ of the outgoing state [cf.\ Eq.\ (\ref{Ewsum}]. The second term ($p=2$, $q=0$) and the third term ($p=0$, $q=2$) are separable states. Only the first term ($p=q=1$) gives a nonzero contribution,\footnote{In the case of spin-independent scattering, when $\gamma=2^{-1/2}\sigma_{y}$, the state $|\Phi\rangle$ simplifies to $|\Phi\rangle=-i\,2^{-1/2}(b^{\dagger}_{L,1}b^{\dagger}_{R,2}-b^{\dagger}_{L,2}b^{\dagger}_{R,1})|0\rangle\equiv -i\,2^{-1/2}(|\!\uparrow\rangle_{e}|\!\downarrow\rangle_{e}-|\!\downarrow\rangle_{e}|\!\uparrow\rangle_{e})$, which is the spin singlet state of Eq.\ (\ref{Belldef}). The electron-hole Bell pair $2^{-1/2}(|\!\uparrow\rangle_{h}|\!\uparrow\rangle_{e}$ +$ |\!\downarrow\rangle_{h}|\!\downarrow\rangle_{e})$ of Eq.\ (\ref{0mapsto}) is obtained by a canonical transformation from electron to hole degrees of freedom at the left of the tunnel barrier.}
\begin{subequations}
\label{Epartresult}
\begin{eqnarray}
&&{\cal E}_{\rm part}=\frac{eVt_{\rm det}}{h}\,w\,{\cal E}(|\Phi\rangle),\label{Epartresulta}\\
&&|\Phi\rangle=\bm{b}_{L}^{\dagger}\gamma \bm{b}_{R}^{\dagger}|0\rangle,\;\;
\gamma=w^{-1/2}\,r\sigma_{y}t^{T},\label{Epartresultb}\\
&&w={\rm Tr}\,(r\sigma_{y}t^{T})(r\sigma_{y}t^{T})^{\dagger}=\tau_{1}+\tau_{2}-2\tau_{1}\tau_{2}. \label{Epartresultc}
\end{eqnarray}
\end{subequations}
The factor $eVt_{\rm det}/h$ counts the number of independently contributing energy intervals.

The degree of entanglement of the two-particle pure state $|\Phi\rangle$ is quantified by the concurrence. Substituting Eq.\ (\ref{Epartresult}) into Eq.\ (\ref{Cdef}) one arrives at the result (\ref{Cdef4}). The result (\ref{ECrelation3}) for the entanglement entropy follows from Eq.\ (\ref{ECrelation}).

\section{Entanglement production at finite temperature}
\label{app_finiteT}

In this appendix we calculate the free-fermion entanglement produced at a tunnel barrier for the case that the electron sources are at finite temperature, generalizing the zero-temperature calculation of Ref.\ \cite{Bee03}. This calculation not been previously published.

We refer to the beam splitter geometry of Fig.\ \ref{fig_beamsplitter}. The incoming state $\rho_{\rm in}$ produced by the sources at an elevated temperature and/or voltage is scattered via a beam splitter into an outgoing state $\rho_{\rm out}$, which is absorbed by detectors at zero temperature and voltage. (The detectors should be kept at a lower temperature than the sources, in order to be able to distinguish incoming from outgoing states.) For electron sources in local thermal equilibrium, the density matrix has the form
\begin{equation}
\rho_{\rm in}=\prod_{n}\prod_{E}\biggl[\bigl(1-f_{n}(E)\bigr)|0\rangle\langle 0|+f_{n}(E)a_{n}^{\dagger}(E)|0\rangle\langle 0|a_{n}^{\vphantom{\dagger}}(E)\biggr] .\label{rhoinT}
\end{equation}
Here $a_{n}(E)$ is the fermion annihilation operator for an incoming channel $n$ at energy $E$ and $f_{n}(E)$ is the Fermi-Dirac distribution. (Different channels may have a different temperature and/or voltage, which is why we speak of local rather than global thermal equilibrium.) The state $|0\rangle$ is the vacuum state (such that $a_{n}|0\rangle=0$).

The annihilation operators $b_{n}(E)$ for the outgoing channels are related to those for the incoming channels by the unitary scattering matrix of the tunnel barrier,
\begin{equation}
b_{n}(E)=\sum_{m}S_{nm}(E)a_{m}(E).\label{barelation}
\end{equation}
Substitution of Eq.\ (\ref{barelation}) into Eq.\ (\ref{rhoinT}) gives the density matrix of the outgoing state in a given energy interval $\delta E$,
\begin{equation}
\rho_{\rm out}=\prod_{n}\biggl[(1-f_{n})|0\rangle\langle 0|+f_{n}c_{n}^{\dagger}|0\rangle\langle 0|c_{n}^{\vphantom{\dagger}}\biggr] ,\;\; c_{n}^{\dagger}=\sum_{m}b_{m}^{\dagger}S_{mn}.\label{rhooutT}
\end{equation}

We consider the case of a single spin-degenerate mode, so $n=1,2$ labels spin up and down from source $A$ and $n=3,4$ labels spin up and down from source $B$. We denote $f_{1}=f_{2}\equiv f_{A}$ and $f_{3}=f_{4}\equiv f_{B}$. The sources are taken at the same temperature $T$ with a voltage difference $V$,
\begin{eqnarray}
&&f_{A}(E)=\frac{1}{\exp[(E-E_{F}-eV)/k_{B}T]+1},\nonumber\\
&&f_{B}(E)=\frac{1}{\exp[(E-E_{F})/k_{B}T]+1}.\label{fAfBdef}
\end{eqnarray}
The Fermi energy at zero voltage is denoted $E_{F}$.

To determine the entanglement production, constrained by particle conservation, we need to project the density matrix $\rho_{\rm out}$ onto sectors ${\mathcal N}_{pq}$ of Fock space with a well-defined particle number $p,q\in\{0,1,2\}$ at detector $A$ and $B$ in each energy interval $\delta E$. In accordance with Eq.\ (\ref{Ewsum}), the entanglement production ${\mathcal E}_{\rm part}$ is the weighted sum of the entanglement in the individual sectors. The only sector with nonzero entanglement is ${\mathcal N}_{11}$, with one particle at detector $A$ and one particle at detector $B$. In all other sectors (with $pq=0$, mod $2$) the projection $\rho_{pq}$ is separable in the two degrees of freedom at detectors $A$ and $B$.

The projected density matrix $\rho_{11}$ is constructed from Eq.\ (\ref{rhooutT}) by summing over the six nonequivalent ways to partition the indices $1,2,3,4$ into two filled channels $n_{1},n_{2}$ and two empty channels $n_{3},n_{4}$,
\begin{widetext}
\begin{eqnarray}
w_{11}\rho_{11}&=&\sum_{\rm partitions}f _{n_{1}}f _{n_{2}}(1-f _{n_{3}})(1-f _{n_{4}})
\bm{b}_{A}^{\dagger}M(n_{1},n_{2})\bm{b}_{B}^{\dagger}|0\rangle\langle 0|\bm{b}_{B}M^{\dagger}(n_{1},n_{2})\bm{b}_{A},\label{rho11}\\
w_{11}&=&\sum_{\rm partitions}f _{n_{1}}f _{n_{2}}(1-f _{n_{3}})(1-f _{n_{4}}){\rm Tr}\,M(n_{1},n_{2})M^{\dagger}(n_{1},n_{2}), \label{w11}\\
M(n_{1},n_{2})&=&
\left(\begin{array}{cc}
S_{1n_{1}}&S_{1n_{2}}\\S_{2n_{1}}&S_{2n_{2}}
\end{array}\right)
\sigma_{y}
\left(\begin{array}{cc}
S_{3n_{1}}&S_{3n_{2}}\\S_{4n_{1}}&S_{4n_{2}}
\end{array}\right)^{T}.\label{Mdef}
\end{eqnarray}
\end{widetext}
We have defined the vectors $\bm{b}_{A}=(b_{1},b_{2})$, $\bm{b}_{B}=(b_{3},b_{4})$ of annihilation operators at detectors $A$ and $B$, respectively.

The general solution simplifies considerably if the scattering matrix $S$ is spin-independent. It then has the block diagonal form
\begin{equation}
S=\begin{pmatrix}
e^{i(\phi{\vphantom'}+\phi')}\openone_{2}\sqrt{1-\tau}&e^{i(\phi{\vphantom'}+\psi')}\openone_{2}\sqrt{\tau}\\
e^{i(\psi{\vphantom'}+\phi')}\openone_{2}\sqrt{\tau}&-e^{i(\psi{\vphantom'}+\psi')}\openone_{2}\sqrt{1-\tau}
\end{pmatrix},\label{Sblock}
\end{equation}
with $\openone_{2}$ the $2\times 2$ unit matrix. We have parameterized the transmission and reflection amplitudes in terms of $4$ phases $\phi{\vphantom'},\phi',\psi{\vphantom'},\psi'$, and the transmission probability $\tau\in[0,1]$.

The projected density matrix (\ref{rho11}) becomes
\begin{eqnarray}
w_{11}\rho_{11}&=&A_{1}\left(|\uparrow\uparrow\rangle\langle\uparrow\uparrow|+|\downarrow\downarrow\rangle\langle\downarrow\downarrow|\right)+A_{2}\,|\Psi_{\rm Bell}\rangle\langle\Psi_{\rm Bell}|\nonumber\\
&&\mbox{}+A_{3}\left(|\Psi_{\gamma}\rangle\langle\Psi_{\gamma}|+\sigma_{x}\otimes\sigma_{x}|\Psi_{\gamma}\rangle\langle\Psi_{\gamma}|\sigma_{x}\otimes\sigma_{x}\right),\nonumber\\
&&\label{rho11nospin}\\
w_{11}&=&4f_{A}(1-f_{A})f_{B}(1-f_{B})+2(f_{A}-f_{B})^{2}\tau(1-\tau),\nonumber\\
&&\label{w11nospin}\\
A_{1}&=&f_{A}(1-f_{A})f_{B}(1-f_{B}),\label{Adef}\\
A_{2}&=&2[f_{A}^{2}(1-f_{B})^{2}+f_{B}^{2}(1-f_{A})^{2}]\tau(1-\tau),\label{Bdef}\\
A_{3}&=&f_{A}(1-f_{A})f_{B}(1-f_{B})[1-2\tau(1-\tau)].\label{A3def}
\end{eqnarray}
We denote the spin channels by $\uparrow,\downarrow$, so that, for example, $b_{1}^{\dagger}b_{4}^{\dagger}|0\rangle=|\uparrow\downarrow\rangle$. In Eq.\ (\ref{rho11nospin}), the first term $\propto A_{1}$ contributes two separable states, the second term $\propto A_{2}$ contributes the maximally entangled Bell state (\ref{Belldef}), and the third term $\propto A_{3}$ contributes a mixture of the partially entangled state
\begin{equation}
|\Psi_{\gamma}\rangle=[1-2\tau(1-\tau)]^{-1/2}[(1-\tau)|\uparrow\downarrow\rangle+\tau|\downarrow\uparrow\rangle]\label{Psigamma}
\end{equation}
and its spin-inverse.

With some more algebra the density matrix (\ref{rho11nospin}) reduces to the Werner state (\ref{rhoWerner}),
\begin{eqnarray}
\rho_{11}&=&\tfrac{1}{4}(1-\xi)\openone_{4}+\xi|\Psi_{\rm Bell}\rangle\langle\Psi_{\rm Bell}|, \label{rho11Werner}\\
\xi&=&(2/w_{11})(f_{A}-f_{B})^{2}\tau(1-\tau)\nonumber\\
&=&1-\frac{1}{1+2\tau(1-\tau)\sinh^{2}u}.\label{rho11xi}
\end{eqnarray}
We have abbreviated $u=eV/2k_{B}T$. Notice that the energy dependence introduced by the Fermi-Dirac distribution drops out of the final expression for $\xi$. The concurrence of the Werner state is given by Eq.\ (\ref{CWerner}). Substitution into Eqs.\ (\ref{ECrelation2}) and (\ref{Ewsum}) gives the entanglement production $\delta {\cal E}_{\rm part}$ in the energy interval $\delta E$ under consideration,
\begin{equation}
\delta{\cal E}_{\rm part}=w_{11}{\cal F}\left(\tfrac{1}{2}+\tfrac{1}{2}\sqrt{1-\tfrac{1}{4}(3\xi-1)^{2}}\right)\theta(\xi-\tfrac{1}{3}),\label{Cdef3}
\end{equation}
with $\theta(x)$ the unit step function [$\theta(x)=0$ if $x<0$, $\theta(x)=1$ if $x>0$].

We assume that the transmission probability is approximately energy independent on the scale of $k_{B}T$. Then the entire energy dependence of $\delta{\cal E}_{\rm part}$ lies in the weight factor $w_{11}$. Integration over energy gives the total entanglement production
\begin{eqnarray}
{\cal E}_{\rm part}&=&\frac{eVt_{\rm det}}{h}{\cal F}\left(\tfrac{1}{2}+\tfrac{1}{2}\sqrt{1-\tfrac{1}{4}(3\xi-1)^{2}}\right)\nonumber\\
&&\mbox{}\times\theta(\xi-\tfrac{1}{3})\frac{\cosh u-u^{-1}\sinh u}{(1-\xi)\sinh^{3}u}.\label{Eparttotal}
\end{eqnarray}
This function is plotted in Fig.\ \ref{fig_Epartplot}.

The maximal entanglement production rate of $eV/2h$ bits per second is reached for $T=0$, $\tau=1/2$. At nonzero temperature the entanglement production decreases, vanishing $\propto(T_{c}-T)^{2}\ln(T_{c}-T)^{-1}$ at the critical temperature $T_{c}$ determined by Eq.\ (\ref{Tceq}).

\section{Bell inequality with noise correlators}
\label{app_c}

\subsection{Tunneling regime}
\label{app_ctunneling}

Following Ref.\ \cite{Bee03}, we calculate the maximal violation of the CHSH inequality (\ref{CHSHdef}) that follows using the expression (\ref{CabC0relation}) for the spin-spin correlator $C_{\bm{ab}}$ in terms of spin-resolved noise correlators. 

At low temperatures ($k_{B}T\ll eV$) the zero-frequency noise correlator $C_{ij}(0)$ has the general expression \cite{Les89,But90,Bla00}
\begin{equation}
C_{ij}=-(e^{3}V/h)|(rt^{\dagger})_{ij}|^{2},\label{Cijrtrelation}
\end{equation}
in terms of the $2\times 2$ reflection and transmission matrices $r,t$ of the tunnel barrier [cf.\ Eq.\ (\ref{inputoutput})]. It is convenient to parameterize these matrices in terms of the transmission eigenvalues $\tau_{1},\tau_{2}$, by
\begin{equation}
r=U_{L}\begin{pmatrix}
\sqrt{1-\tau_{1}}&0\\0&\sqrt{1-\tau_{2}}
\end{pmatrix}U_{0},\;\;
t=U_{R}\begin{pmatrix}
\sqrt{\tau_{1}}&0\\0&\sqrt{\tau_{2}}
\end{pmatrix}U_{0},\label{rtparam}
\end{equation}
with $2\times 2$ unitary matrices $U_{L},U_{R},U_{0}$. [In the presence of time-reversal symmetry one has $U_{0}=U_{L}^{T}$, but this constraint is irrelevant because anyway $U_{0}$ drops out of Eq.\ (\ref{Cijrtrelation}).]

The spin-spin correlator $C_{\bm{ab}}$ depends on the directions $\bm{a}$ and $\bm{b}$ along which the spin is measured at the left and right of the tunnel barrier. Combination of Eqs.\ (\ref{CabC0relation}) and (\ref{Cijrtrelation}) gives
\begin{equation}
C_{\bm{ab}}=\frac{{\rm Tr}\,(\bm{a}\cdot\bm{\sigma})rt^{\dagger}(\bm{b}\cdot\bm{\sigma})tr^{\dagger}}{{\rm Tr}\,rt^{\dagger}tr^{\dagger}}.\label{Cabrtrelation}
\end{equation}
Equivalently, one can measure the spins along the fixed $z$-axis after having rotated them by unitary matrices $U_{\bm a}$ and $U_{\bm b}$, such that
\begin{equation}
U_{\bm a}^{\dagger}\sigma_{z}U_{\bm a}^{\vphantom{\dagger}}=\bm{a}\cdot\bm{\sigma},\;\;
U_{\bm b}^{\dagger}\sigma_{z}U_{\bm b}^{\vphantom{\dagger}}=\bm{b}\cdot\bm{\sigma}.
\label{Uarelation}
\end{equation}
In the parameterization (\ref{rtparam}) one finds
\begin{eqnarray}
C_{\bm{ab}}&=&(1-2|\tilde{U}_{\bm{a},11}|^{2})(1-2|\tilde{U}_{\bm{b},11}|^{2})\nonumber\\
&&\mbox{}+4\kappa{\cal C}\,{\rm Re}\,\tilde{U}^{\vphantom{\ast}}_{\bm{a},11}\tilde{U}^{\ast}
_{\bm{b},11}\tilde{U}^{\ast}_{\bm{a},12}\tilde{U}^{\vphantom{\ast}}_{\bm{b},12},
\label{EabCrelation}\\
\tilde{U}_{\bm{a}}&=&U_{\bm{a}}U_{L},\;\;\tilde{U}_{\bm{b}}=U_{\bm{b}}U_{R},\label{UUtilde}\\
\kappa&=&1+\frac{(\tau_{1}-\tau_{2})^{2}}{\tau_{1}(1-\tau_{1})+\tau_{2}(1-\tau_{2})}.\label{kappadef}
\end{eqnarray}
The quantity ${\cal C}$ in Eq.\ (\ref{EabCrelation}) is the concurrence (\ref{Cdef4}) of the electron-hole pair.

By varying the unit vectors $\bm{a},\bm{b},\bm{a'},\bm{b'}$ (or, equivalently, the unitary matrices $\tilde{U}_{\bm{a}}$, $\tilde{U}_{\bm{b}}$, $\tilde{U}_{\bm{a'}}$, $\tilde{U}_{\bm{b'}}$) in the CHSH inequality (\ref{CHSHdef}) one finds the maximum violation
\begin{equation}
{\cal B}_{\rm max}=2\sqrt{1+\kappa^{2}{\cal C}^{2}}\approx 2\sqrt{1+{\cal C}^{2}}\;\;{\rm if}\;\;\tau_{1},\tau_{2}\ll 1.\label{Bmaxresult}
\end{equation}
This confirms the expected relation (\ref{BmaxCrelation}) between the concurrence and the Bell parameter in the tunneling regime $\tau_{1},\tau_{2}\ll 1$.

\subsection{Beyond the tunneling regime}
\label{app_cnotunneling}

Following Ref.\ \cite{Bee04a}, we calculate the maximal violation of the CHSH inequality (\ref{CHSHdef}) that follows using the expression (\ref{CabC0relation2}) for the spin-spin correlator $C_{\bm{ab}}$. This expression contains both the noise correlators and the mean currents and does not require the tunneling approximation. Referring to the beam splitter geometry of Fig.\ \ref{fig_beamsplitter}, only the current from the beam splitter to the detectors should be included in the mean current, not the current from the sources to the beam splitter\footnote{This distinction between outgoing and incoming particles is irrelevant in the tunneling regime at zero temperature, because the non-fluctuating incoming current does not contribute to the noise correlator $C_{ij}(0)$. In an electronic circuit the outgoing current can be measured separately from the incoming current in a four-terminal geometry with grounded detectors at zero temperature.}. 

The mean currents into the detectors are given by
\begin{equation}
\langle I_{A,i}\rangle=\frac{e^{2}V}{h}(rr^{\dagger})_{ii},\;\;
\langle I_{B,i}\rangle=\frac{e^{2}V}{h}(tt^{\dagger})_{ii}.
\label{IAIBmean}
\end{equation}
Combination of Eqs.\ (\ref{CabC0relation2}), (\ref{Cijrtrelation}), and (\ref{IAIBmean}) gives
\begin{equation}
C_{\bm{ab}}=\frac{[{\rm Tr}\,(\bm{a}\cdot\bm{\sigma})rr^{\dagger}][{\rm Tr}\,(\bm{b}\cdot\bm{\sigma})tt^{\dagger}]-{\rm Tr}\,(\bm{a}\cdot\bm{\sigma})rt^{\dagger}(\bm{b}\cdot\bm{\sigma})tr^{\dagger}}{({\rm Tr}\,rr^{\dagger})({\rm Tr}\,tt^{\dagger})-{\rm Tr}\,rt^{\dagger}tr^{\dagger}}.\label{Cabrtrelation2}
\end{equation}

Substitution of Eqs.\ (\ref{rtparam}) and (\ref{Uarelation}) leads to
\begin{eqnarray}
C_{\bm{ab}}&=&-(1-2|\tilde{U}_{\bm{a},11}|^{2})(1-2|\tilde{U}_{\bm{b},11}|^{2})\nonumber\\
&&\mbox{}-4{\cal C}\,{\rm Re}\,\tilde{U}^{\vphantom{\ast}}_{\bm{a},11}\tilde{U}^{\ast}
_{\bm{b},11}\tilde{U}^{\ast}_{\bm{a},12}\tilde{U}^{\vphantom{\ast}}_{\bm{b},12}.
\label{EabCrelation2}
\end{eqnarray}
Apart from an overall minus sign, Eq.\ (\ref{EabCrelation2}) is the same as Eq.\ (\ref{EabCrelation}) without the spurious factor $\kappa$ multiplying the concurrence. The maximal violation of the Bell inequality becomes
\begin{equation}
{\cal B}_{\rm max}=2\sqrt{1+{\cal C}^{2}}.\label{Bmaxresult2}
\end{equation}
The relation (\ref{BmaxCrelation}) between concurrence and Bell parameter is now satisfied beyond the tunneling regime.

\end{document}